\documentclass[floats,floatfix,aps,showpacs,twocolumn,pre,superscriptaddress]{revtex4}
\usepackage{amsfonts}
\usepackage{amssymb}
\usepackage{amsmath}
\usepackage{dcolumn}
\usepackage{bm}
\usepackage[]{graphicx}
\usepackage{mathrsfs}  			
\usepackage{bm}

\newcommand{\Bra}[1]{\left( #1 \right)}				%( #1 )
\newcommand{\BRA}[1]{\left[ #1 \right]}				%[ #1 ]
\newcommand{\ket}[1]{| #1\rangle}			%| #1>
\newcommand{\bra}[1]{\langle #1|}			% < #1| 
\newcommand{\bracket}[2]{\langle #1 | #2 \rangle} 	%<#1|#2>
\newcommand{\ketbra}[2]{| #1 \rangle\!  \langle #2 |} 	%|#1><#2|

\newcommand{\Con}[0]{\mathrm{Con}_\ell^{(M)}}
\newcommand{\AbsU}[0]{\hat{U}^{\mathsf{o}}}
\newcommand{\eps}[0]{\varepsilon}
\newcommand{\BCH}[0]{\hat{H}_\mathrm{eff}^{(M)}}
\newcommand{\Heff}[0]{\hat{H}_\mathrm{eff}}

\begin{document}	

\title{Origin of the enhancement of tunneling probability in the nearly integrable system}
\author{Yasutaka Hanada}
%\email{hanada-yasutaka@ed.tmu.ac.jp}
\affiliation{Department of Physics, Tokyo Metropolitan University, Tokyo 192-0397, Japan}
\author{Akira Shudo}
\affiliation{Department of Physics, Tokyo Metropolitan University, Tokyo 192-0397, Japan}
\author{Kensuke S. Ikeda}
\affiliation{Department of Physics, Ritsumeikan University, Kusatsu, Shiga, 525-0577, Japan}

\date{\today}
\begin{abstract}
%%% ShShSh(2014.9.23) 
The enhancement of tunneling probability in the nearly integrable system 
is closely examined, focusing on tunneling splittings plotted as a function 
of the inverse of the Planck's constant. 
On the basis of the analysis using the absorber which efficiently suppresses 
the coupling creating spikes in the plot, we found 
that the splitting curve should be viewed as 
the staircase-shaped skeleton accompanied by spikes. 
We further introduce renormalized integrable Hamiltonians, 
and explore the origin of such a staircase structure by 
investigating the nature of eigenfunctions closely. 
It is found that the origin of the staircase structure could trace back to 
the anomalous structure in tunneling tail which 
manifests itself in the representation using renormalized action bases.
This also explains the reason why the staircase does not appear in the completely integrable system. 
\end{abstract}
\maketitle

\section{Introduction}\label{sec:introduction}

The tunneling effect is peculiar to quantum mechanics and
no counterparts exist in classical mechanics.  
Quantum tunneling plays a role and actually manifests in various situations
ranging from atomic and molecular physics to 
various phenomena in condensed phases. 

In most cases, incorporating the tunneling effect into each case is made by 
using the system with a single degree of freedom. 
This is justified and certainly provides a good description if the tunneling penetration 
proceeds in only one direction, but this is not the case when the system has 
multi degrees of freedom. 

The most important qualitative difference between one- and multi-dimensional 
systems would be that classical particles are confined not only 
by the energy barrier, but also by the {\it dynamical barrier}. 
The latter is formed when additional constants of motion, either globally or locally,  
exist besides the energy. 
What is more crucial is the fact that 
generic multi-dimensional systems are no more completely integrable 
and chaos appears in the underlying classical dynamics, so one must take into account 
new aspects of quantum tunneling absent in completely integrable systems \cite{creagh_1998,TunnelBook}. 

The role of classical chaos in quantum tunneling has first been discussed 
in the observation of the wave packet dynamics \cite{lin_1990}, and then clearly 
recognized in the behavior of the tunneling splitting of eigenenergies 
\cite{bohigas_1990,tomsovic_1994,roncaglia_1994}. 
To understand why chaos could play a role in the tunneling process, 
it suffices to suppose the states forming a doublet, which is a complete analog of 
the doublet appearing in the system with one-dimensional symmetric double well potential. 
It is important to note, however, that the doublet in multi-dimensional systems are supported 
by symmetric regular tori in phase space and chaos exists in between. 
As one varies an external parameter of the system, it can happen that 
states forming the doublet and a state supported by the chaotic region 
come close to each other in the energy space and form avoided crossing. 
Within the interaction regime, the energy splitting between 
the doublet becomes large through couplings with the chaotic state, 
meaning that the tunneling amplitude between one torus to the other is enhanced. 
{\it Chaos-assisted tunneling} (CAT) occurs in this way 
\cite{bohigas_1990,tomsovic_1994}.  
A similar enhancement is known to take place if 
the doublet is bridged by nonlinear resonances. 
Nonlinear resonances are also 
important ingredients in multi-dimensional systems, 
and the latter mechanism is called {\it resonance-assisted tunneling} (RAT) 
\cite{bonci_1998,brodier,eltschka_2005,mouchet_2006,schlagheck_2011,laeck2010}. 

In order to go beyond qualitative explanations 
and to obtain more direct evidence for the connection between 
chaos (and/or nonlinear resonances) 
%and the enhancement of quantum tunneling, 
%convincing understanding for newly discovered types of tunneling, 
semiclassical (WKB) analyses are desired, especially based on 
the complex classical dynamics since quantum tunneling is a classically forbidden process. 

For the system with 
one degree of freedom, there indeed exists a standard approach that has 
been established already \cite{coleman_1977}. The {\it instanton} is the name of a complex orbit which 
conveys the tunneling amplitude running along the imaginary time axis, and the formula 
representing the tunneling splitting in the symmetric double well mode is expressed as 
\begin{equation}\label{eq:splitting_fomula}
  \Delta E \sim \alpha\hbar e^{-S/\hbar}, 
\end{equation}
where and classical ingredients $\alpha$ and $S$
can be deduced in the instanton calculation \cite{coleman_1977}.

On the other hand, in multi-dimensional cases, full semiclassical analyses using 
the complex classical dynamics could so far be applied only to the time domain 
\cite{shudo_1998,shudo_2009} and have not been even 
formulated except for completely integrable situations in the energy domain 
\cite{wilkinson_1986,creagh_1994,deunff_2010,deunff_2013}. 
The enhancement of tunneling could therefore 
be well accounted for in terms of 
the complex dynamics \cite{shudo_2012}, but fully convincing 
semiclassical understanding for the energy domain is still lacking.

The aim of this article is to explore the origin of the enhancement of 
tunneling probability observed in the energy domain. 
The tunneling probability in the energy domain is often measured 
by tracking the energy splitting or the properly defined tunneling rate  
as a function of $1/h$. The enhancement is typically 
observed as plateaus accompanied by spikes due to energy resonance \cite{roncaglia_1994,brodier}. 
Characteristics observed there are understood within a framework of RAT, and  
in particular,  plateaus could be interpreted as a kind of phenomena that  
might be called quantum overlapping resonances; a bunching of spikes, 
each of which is associated with an individual quantum resonance,  
turns out to create plateaus, or a persistent long-range interaction of each resonance with 
other states \cite{laeck2010,schlagheck_2011}. 

However, one should recall that the enhancement occurs 
even when the Planck's constant 
is not small enough to resolve the chaotic components or nonlinear resonance islands. 
This rather paradoxical behavior has been already observed in several models 
\cite{shudo_2012,ikeda_2013}, 
but its origin in such a slightly perturbed regime 
has never been seriously investigated to the authors' knowledge. 

We here take a close look at the nature of the enhancement of the tunneling 
probability in such nearly integrable regimes by introducing techniques such 
as absorbing the individual states involved in avoided crossings  and decomposing 
the eigenstates into proper integrable bases, as explained in detail below. 
We especially focus not only on the behavior of the energy levels but also on 
the nature of eigenfunctions to elucidate which components 
are mostly responsible for the plateau structure formed in the energy splitting 
vs $1/h$ plot. 

In conjunction with this, 
we shall stress the importance of observing wavefunctions in the whole range 
because there are various ways to define the $\lq\lq$tunneling probability", 
and the nature of tunneling may look different depending on how it is defined.
Here, for the closed system (standard map) 
the splitting of energy levels will be adopted to measure the tunneling probability, 
whereas for the open system, like the H\'enon map, the probability in the asymptotic region is naturally introduced, 
and the decay rate for the absorbed system is sometimes used. 
There would be no legitimate way or one should even say that 
providing a proper definition of the tunneling probability itself 
is an issue to be explored in nonintegrable systems. 
Therefore, one should examine more carefully the tail of wavefunctions 
in the whole range before focusing on the amplitude at a certain specific position.
%otherwise the result may mislead us. 

The present analysis is motivated by a recent work in which the mechanism of 
the {\it instanton-noninstanton} (I-NI) {\it transition} has closely been studied in terms of 
quantum perturbation theory 
%based on the Baker-Campbell-Hausdorff (BCH) expansion 
\cite{ikeda_2013}, 
and so spirits and tools for analyses are overall common.  The term instanton-noninstanton (I-NI)
is named after 
the first transition at which the deviation from the instanton prediction starts \cite{ikeda_2013}.

The organization of the paper is as follows: 
In section \ref{sec:model}, we introduce the system studied in this paper, 
and present aspects of the enhancement of the tunneling probability by 
observing the quantum number and $1/h$ dependence of the tunneling probability 
in our model.
In section \ref{sec:staircase}, introducing an absorbing operator, which projects out 
a given set of integrable states, we examine which states are responsible for 
creating spikes typically observed in the splitting curve 
and whether or not the staircase structure of the splitting curve appears
as a result of local quantum resonances in the energy space. 
In section \ref{sec:eigenstate}, we investigate the nature of eigenstates to clarify the mechanism 
of the enhancement by focusing on the local probability amplitude of eigenfunctions and 
the contribution spectrum introduced in \cite{ikeda_2013}. 
In section \ref{sec:rat_integ}, on the basis of analyses made in section \ref{sec:eigenstate} 
we claim that an essential difference of the splitting curve exists 
between integrable and nonintegrable systems.
In the final section, we summarize and provide outlook especially toward our forthcoming papers. 

\section{Enhancement of tunneling probability}
\label{sec:model}

\begin{figure}[t]
	\center 
	\includegraphics[width=0.45\textwidth]{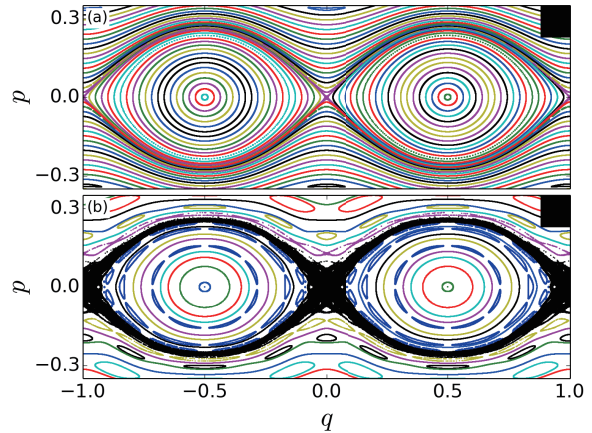}
	\caption{\label{fig:phase_space}
	(Color online)
	Classical phase space for the symmetrized standard map $f$ with (a) $\tau=2/3$ and (b) $\tau=1$. 
	There are no visible resonance chains for $\tau=2/3$ in the inner tours region
	while chaotic regions around the unstable fixed point at $(q,p)=(0,0)$
	and some resonance chains ($1\colon8$, $1\colon10$, $1\colon12$,$\cdots$)
	become visible for $\tau=1$. They are shown in dark blue. 
	The black box put in the upper right corner indicates the size of the effective Planck's constant 
	for $h=1/70$.
}
\end{figure}

We consider a quantum system described by the evolution operator
in a symmetrized form: 
\begin{equation}\label{eq:qmap}
  \hat{U}=e^{-\frac{i}{\hbar}V(\hat{q})\tau/2}
  e^{-\frac{i}{\hbar}T(\hat{p})\tau}
  e^{-\frac{i}{\hbar}V(\hat{q})\tau/2}. 
\end{equation}
The corresponding classical dynamics is given as the symplectic map
$f:=f_V(\frac{\tau}{2})\circ f_T(\tau)\circ f_V(\frac{\tau}{2})$
where 
$f_V(\tau): (q,p)\mapsto(q,p+\tau V'(q))$ and $f_T(\tau): (q,p)\mapsto(q + \tau T'(p),p )$ 
are trivial symplectic maps. Here the prime stands for the derivative of the function.
The classical map $f$ 
corresponds to discretization of the continuous Hamiltonian flow 
for  $H(q,p) = T(p)+V(q)$ up to the second order of the discrete time step $\tau$.
Thus, the map $f$ 
has the integrable (continuous) limit $\tau\to0$, 
and much the same is true on the quantum map $(\ref{eq:qmap})$.
Hereafter we take the potential function as 
$T(p)=p^2/2$ and $V(q)=(k/4\pi^2)\cos(2\pi q)$
where $k$ is the strength of the perturbation. 
%If $\tau$ equals to 1, 
After rescaling as $p\mapsto p/\tau$ and $k\tau^2=\varepsilon$,
the classical map $f$ turns out to be the symmetrized standard map \cite{chirikov_1969},
and the time evolution by the unitary operator $\hat{U}$ 
can be interpreted as a 
single period evolution of a $\delta$-functional periodically forcing Hamiltonian 
with a period $\tau$. 

In the continuous limit $\tau\to0$, the closed area surrounded by 
the separatrix is given by $S=\sqrt{k}(2/\pi)^2$.
In the following argument, we focus especially on the nearly integrable 
regime and a proper integrable limit will play an important role as a reference. 
In most of situations, nonlinearity is controlled by changing the parameter $\tau$, 
keeping the parameter fixed as $k=k_0\equiv 0.7458$.

Figure \ref{fig:phase_space} displays classical phase space for typical nearly integrable parameter regions. 
In the case of $\tau = 2/3$, 
classical phase space is predominantly covered by 
regular regions and nonlinear resonance chains are not 
visible in this scale. 
For $\tau=1$, small chaotic regions emerge around 
an unstable fixed point at $(q,p)=(0,0)$, and Poincar\'e-Birkhoff chains 
induced by nonlinear resonances become visible. 
Relatively large nonlinear resonances in the inner torus region, 
which represents librational motions in the pendulum Hamiltonian $H$,
are $1:8$, $1:10$, and $1:12$ ones, 
which are marked in dark blue in Fig. \ref{fig:phase_space}(b). 
Below we mainly develop our discussion in the case $\tau=1$,
but essentially the same argument follows for other $\tau$ cases.

We numerically solve the eigenvalue problem for the unitary operator $\hat U$
\begin{align}\label{eq:eigen}
\hat{U}\ket{\Psi_n} = u_n \ket{\Psi_n},
\end{align}
under the periodic boundary condition on the region $(q,p)\in (-1,1]\times (-1/2\tau, 1/2\tau]$.
Let $N$ be the dimension of the Hilbert space 
space, then to achieve the periodic boundary condition the relation $1/2\tau \times 2/\hbar =2\pi N$ 
should hold, which yields the relation 
$h=2/N\tau$.

\begin{figure}
	\center 
	\includegraphics[width=0.42\textwidth]{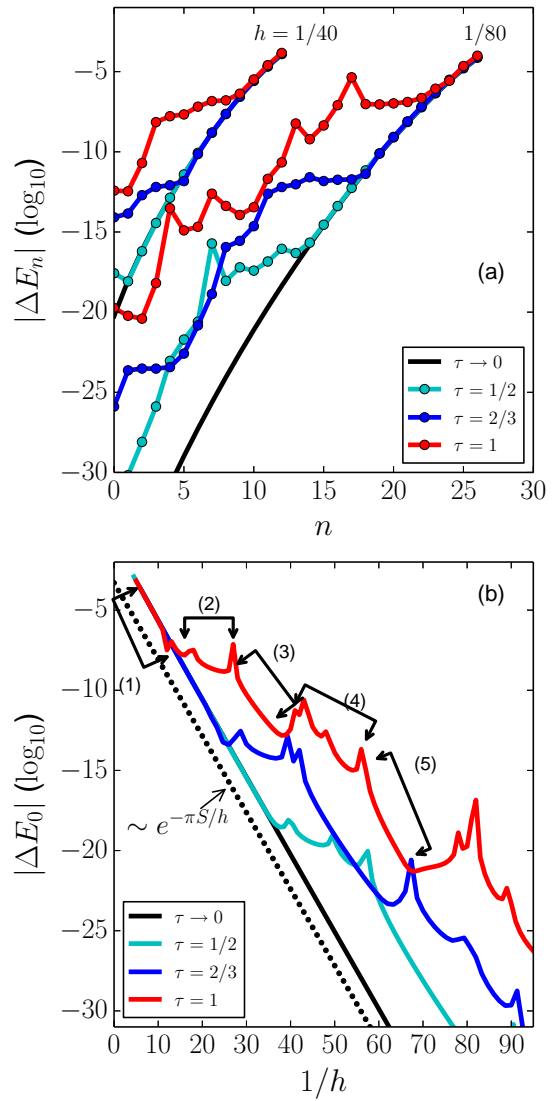}
	\caption{\label{fig:splitting}
	(Color online)
	(a) The tunneling splitting $\Delta E_n$ is plotted as a function of the 
	quantum number $n$, where $n$ is labeled not for an individual state but for a doublet (see the text). 
	The value of $h$ for each curve is put in the figure. 
	The black curves represent the splitting $\Delta E^{(1)}_n$ in the limit ($\tau \to 0$)
	(b) The tunneling splitting $\Delta E_0$ for the lowest doublet 
	$\ket{\Psi_0^{\pm}}$ is shown as a function of $1/h$. 
	The black solid and dotted curves correspond to the splitting $\Delta E^{(1)}_0$ 
	in the limit ($\tau \to 0$)
	and the semiclassical prediction {(\ref{eq:splitting_fomula}}), respectively. 
	In (b), we put the labels (1), (2), $\cdots$, (5) on each characteristic interval: 
	(1) first exponential decay (instanton),
	(2) first plateau,
	(3) second steeply decay,
	(4) second plateau,  and (5) third steeply decay regime, respectively. 
	}
\end{figure}

Here $u_n$ is expressed as $u_n = e^{-iE_n\tau/\hbar}$
where $E_n ~(n =0,1,2,\cdots)$ are quasi-energies, 
and $\ket{\Psi_n}$ denote the corresponding quasi-eigenstates. 
Hereafter we focus on the doublet states in bounded states supported by the inner torus region, each of which is centered at $(q,p)=(\pm 1/2, 0)$ and energy splittings between them. 

Quasi-eigenstates {$\ket{\Psi_n}$} have a symmetry with respect to 
the mirror transformation {$\hat{\Pi}_q:q\mapsto-q$}, and 
we hereafter denote the doublet states associated with this symmetry 
by $\ket{\Psi_n^{\pm}}$ and the corresponding quasi-energies by $E_n^{\pm}$, 
which form quasi degeneracy.  We therefore assign the quantum number $n$ not 
to an individual quasi-eigenstate but to each doublet \cite{saddle_numbering}.
The states $\ket{\Psi_n^{+}}$ and $\ket{\Psi_n^{-}}$ respectively represent 
symmetric and anti-symmetric states. 
Note that we have additional symmetry  with respect to the translation 
{$\hat{T}_1:q\mapsto q+1$}, originating from the periodic boundary condition 
in the $q$-direction. This symmetry does not induce quasi degeneracy in energy, 
but the states belonging to a different translational symmetry class do not 
interact with each other even if they have the same mirror symmetry. 

In the continuous limit $\tau\to0$, 
the eigenvalue equation for the Hamiltonian $H(q,p)=p^2/2+V(q)$ 
is expressed as 
\begin{equation}\label{eq:integ_eigen}
	\hat{H}(\hat{q},\hat{p})\ket{J^{\pm}_n} = E_n^{\pm,(1)}\ket{J^{\pm}_n},
\end{equation}
where eigenstates $\ket{J^{\pm}_n}$ are in the same symmetry class as 
the corresponding $\ket{\Psi_n^{\pm}}$.
Here the quantum number $n$
is, as usual, attached 
in ascending order of eigenvalue $E_n^{\pm,(1)}$, so $\ket{J^{\pm}_n}$ 
represents the ground state doublet, which we will hereafter focus on.

For the later purpose, 
we rearrange the quantum number $n$ for the quasi eigenstates $\ket{\Psi^{\pm}_n}$
such that the overlap $|\bracket{J^{\pm}_n}{\Psi^{\pm}_n}|^2$ is maximal. 
This condition, that is one-to-one correspondence between $\ket{\Psi^{\pm}_n}$ 
and $\ket{J^{\pm}_n}$ is fulfilled for the values of $\tau$ used in the present analysis.

With increase in the value of $1/h$, 
the tunneling probability between $\ket{\Psi_n^{\pm}}$, which 
could be measured by the tunneling splitting $\Delta E_n= E_n^{+} - E_n^{-}$, 
becomes large in several orders of magnitude as compared to
those predicted in the continuous limit.  The latter 
is evaluated as $\Delta E^{(1)}_{n}=E^{+,(1)}_{n}-E^{-,(1)}_{n}$. 
%and well fitted by the instanton calculation (see Fig. \ref{fig:splitting}(b)). 
We notice that the overall behavior does not depend on the value of 
the perturbation strength $\tau$, 
although the Planck's cell can resolve chaotic regions and nonlinear island chains 
in the case of $\tau = 1$, whereas this does not the case at all for $\tau=2/3$. 
%This is surprising because it occurs even when the strength of perturbation is 
%so small that the Planck's cell can resolve neither chaotic regions nor nonlinear island chains 
(see Fig. \ref{fig:phase_space} and Fig. {\ref{fig:husimi}}).

We illustrate such anomalous enhancement of tunneling probability 
in a nearly integrable regime in two ways. 
First, as shown in Fig. \ref{fig:splitting}(a), 
the tunneling splitting $\Delta E_n$ is plotted as a function of the quantum number $n$,
and in Fig. \ref{fig:splitting}(b) the splitting $\Delta E_0$ as a function of $1/h$.  
The latter is known to be a standard plot often used in the study of RAT 
\cite{brodier,eltschka_2005,mouchet_2006,laeck2010,schlagheck_2011}.

In Fig. \ref{fig:splitting}(a) we notice that, in the relatively large $n$ regime,
$\Delta E_n$ 
%in each case, $\tau=1/2$, $\tau=2/3$ and $\tau=1$, 
can be fitted by the lines predicted 
by the formula (\ref{eq:splitting_fomula}), 
implying that they have completely integrable nature in essence. 
On the other hand, as $n$ goes down from exited states to the ground state, 
with a fixed $\hbar$, 
the law described by the formula (\ref{eq:splitting_fomula}) is violated at certain critical 
quantum numbers $n_c$, each of which depends on the value of $h$ \cite{brodier,ikeda_2013}. 

At such a quantum number $n_c$,  the curve for $\Delta E_n$ changes its 
slope and forms the plateau. 
After a certain plateau interval, as typically seen in 
$\tau=1$ and $\tau=2/3$ for $h=1/80$,
the slope 
again becomes large, and then forms the second plateau. 
The emergence of plateaus means the enhancement of the tunneling probability
as compared to the integrable (instanton) prediction.
It is particularly non-trivial and even paradoxical because 
this enhancement is relatively stronger in the lower doublets than higher excited ones. 
Note also that the critical quantum number $n_c$ becomes large with increase 
in the value of $\tau$.  
This sudden departure from integrable tunneling has been pointed out 
in the study of RAT \cite{brodier}, 
and it is called the {\it instanton-noninstanton} (I-NI) transition in 
Ref. \cite{shudo_2012,ikeda_2013}, in which 
the mechanism behind it has been investigated in a different  
perspective.

The I-NI transition is similarly observed in the $\Delta E_n$ vs
$1/h$ plot. 
%The latter is known to be a standard plot often used 
%in the study of RAT \cite{brodier,eltschka_2005,mouchet_2006,laeck2010,schlagheck_2011}.
As shown in Fig. \ref{fig:splitting}(b), the energy splitting
$\Delta E_0$ for the lowest doublet $\ket{\Psi_0^{\pm}}$ exhibits a similar behavior. 
For relatively large values of $h$, $\Delta E_0$ follows the instanton prediction 
Eq. (\ref{eq:splitting_fomula}), but deviates from it at certain values of $h$, each of 
which depends on the value of $\tau$. 
The staircase-like structure formed with plateau and steeply decaying intervals again 
characterize the overall structure. 
For the purpose of illustration, we call each region in the staircase, 
(1) first exponential decay (instanton)
(2) first plateau,
(3) second steeply decay,
(4) second plateau,  and (5) third steeply decay regime, 
respectively (see Fig. \ref{fig:splitting}(b)). 

What is prominent in the latter plot than in the former plot is the appearance of spikes. 
This is because in the former plot Fig. \ref{fig:splitting}(a), 
we could evaluate tunneling splitting only at integer values (quantum numbers),
so may miss spikes even if they exist, 
whereas we can scan $\Delta E_0$ at more numerous values of $1/h$. 

The origin of spikes is a central issue in theory of RAT
\cite{eltschka_2005,laeck2010, schlagheck_2011}, 
in which the effect of nonlinear resonance is incorporated by first constructing 
local integrable pendulum Hamiltonian classically 
and then applying quantum perturbation theory. 
Plotting the energy levels as a function of some parameter,  $k$ for example, 
one can recognize that the mechanism of the enhancement due to RAT is similar to CAT: 
as the parameter is varied, the states forming 
the reference doublet, $\ket{\Psi_0^{\pm}}$ in the present case, 
come close to a third state.  They interact with each other, 
%to create avoided crossings. 
and in the interaction regime the splitting between the reference doublet becomes large, 
resulting in a spike \cite{schlagheck_2011}. 
Note, however, that the staircase structure formed with the plateau and 
steeply decaying regime has never been found at least in the completely 
integrable systems studied so far.

\section{Staircase structure with resonance spikes}
\label{sec:staircase}

\subsection{Resonance spikes and the third states}
\label{subsec:resonance}

\begin{figure}[t]
	\center
	\includegraphics[width=0.45\textwidth]{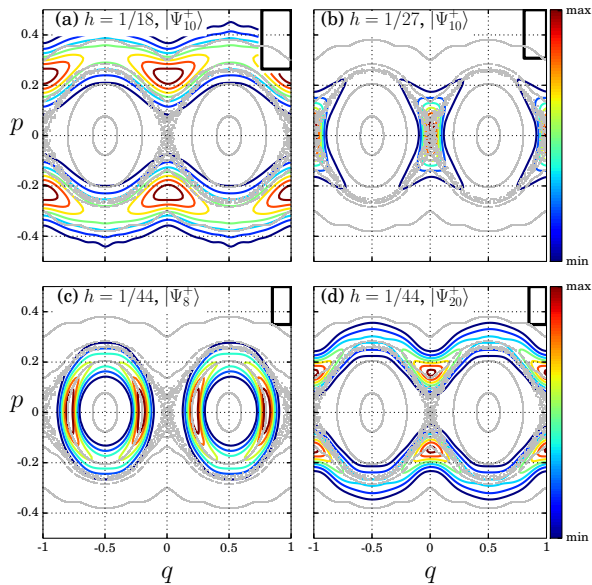}
	\caption{\label{fig:husimi}
	(Color online)
	Husimi representation of the third states for $\tau=1$,
	which resonantly interact with the reference doublet in the first plateau regime:
	(a) $1/h=18$ (at the middle of the first plateau), 
	(b) $h=1/27$ (at the end of first plateau), 
	(c) and (d) $h=1/44$ (at the second plate regime).
	In (a), (c) and (d) we only show $\ket{\Psi_n^{+}}$ states out of each tunneling doublet. 
	In (b), the state $\ket{\Psi_n^{+}}$ is not the one forming a doublet as 
	mentioned in \cite{saddle_numbering}. 
	Upper left box represents the size of effective Planck's cell. 
}
\end{figure}

Each spike observed in Fig. \ref{fig:splitting}(b) appears as a result of
energy resonance between the doublet $\ket{\Psi_0^{\pm}}$ and a certain third state.
The spikes mostly appear in the plateau regime, 
but sometimes they are situated in the steeply decaying regime. 
In Fig. \ref{fig:husimi}, 
we first demonstrate which type of third states are actually involved in the 
creation of spikes. 
In the original framework of the RAT theory, the predicted spikes are associated with 
the states supported by nonlinear resonances in the inner torus region, encircling 
central elliptic fixed points, $(q,p)=(\pm 1/2, 0)$ in the present case. 
However, two of spikes in the first plateau 
appear as a result of resonance with the states associated with 
an outer transversal torus and 
the spike located at the end of the first plateau $h=1/27$ is associated with the state localized 
on the unstable fixed points $(q,p)=(0, 0)$ and $(-1,0)$ (see Fig. \ref{fig:husimi}(b)). 
This is not surprising since 
%the eigenphase $\tau E_n/\hbar$ of Eq. (\ref{eq:eigen}) can satisfy resonance conditions
%$E_n - E_\ell = h m/\tau$ ($n, \ell, m\in\mathbb{Z}$) since the present eigenstates are 
%Floquet states.  
the present eigenstates are Floquet states, 
so the eigenphase $\tau E_n/\hbar$ of Eq. (\ref{eq:eigen}) can satisfy the 
resonance conditions $E_n- E_{\ell} = m h/\tau$ $(n,\ell,m \in {\Bbb Z})$. 
Therefore the quasi-energies of our reference doublet 
can resonate with a state associated with an outer transversal tours. 
Such situations are out of the scope of the theory of RAT, 
but as will closely be discussed in section \ref{sec:eigenstate}, 
the outer torus states play a crucial role in the formation mechanism of the staircase structure. 
%Note that such situations are out of the scope of RAT.
%and thus the prescription developed in the RAT theory would not be applied. 

%On the other hand, 
%in the second plateau, one of the third states generating a spike is associated with 
%a nonlinear resonance in the inner torus region (see Fig. {$\ref{fig:husimi}$}(c)),
%so we may call it a RAT spike. 
%A discussed in subsection \ref{subsec:action}, on the other hand,
%all the spikes observed on plateaus can be attributed simply
%to the quantum resonance caused by the periodic driving.  
%This is just based on the standard time-dependent 
%perturbation argument.  
%We note that the relation of the spikes associated with quantum resonance 
%to those predicted by RAT theory is not clear.

\begin{figure*}[t]
	\centering
	\includegraphics[width=0.4\textwidth]{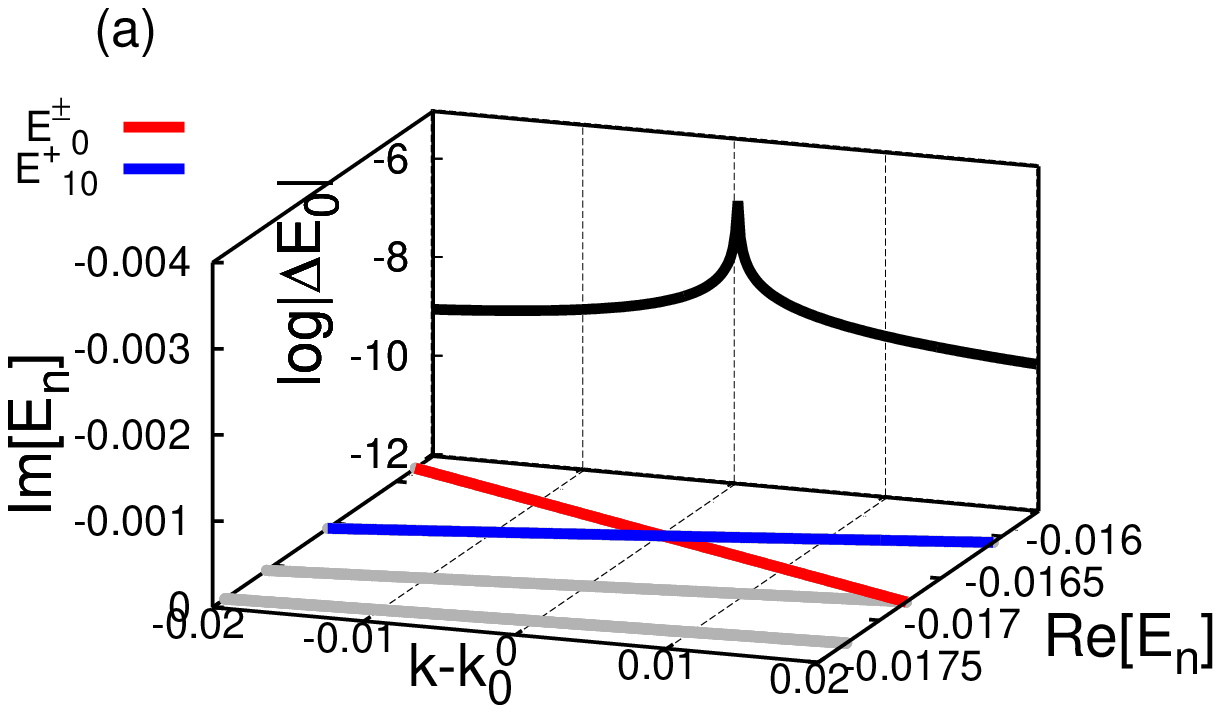}
	\includegraphics[width=0.4\textwidth]{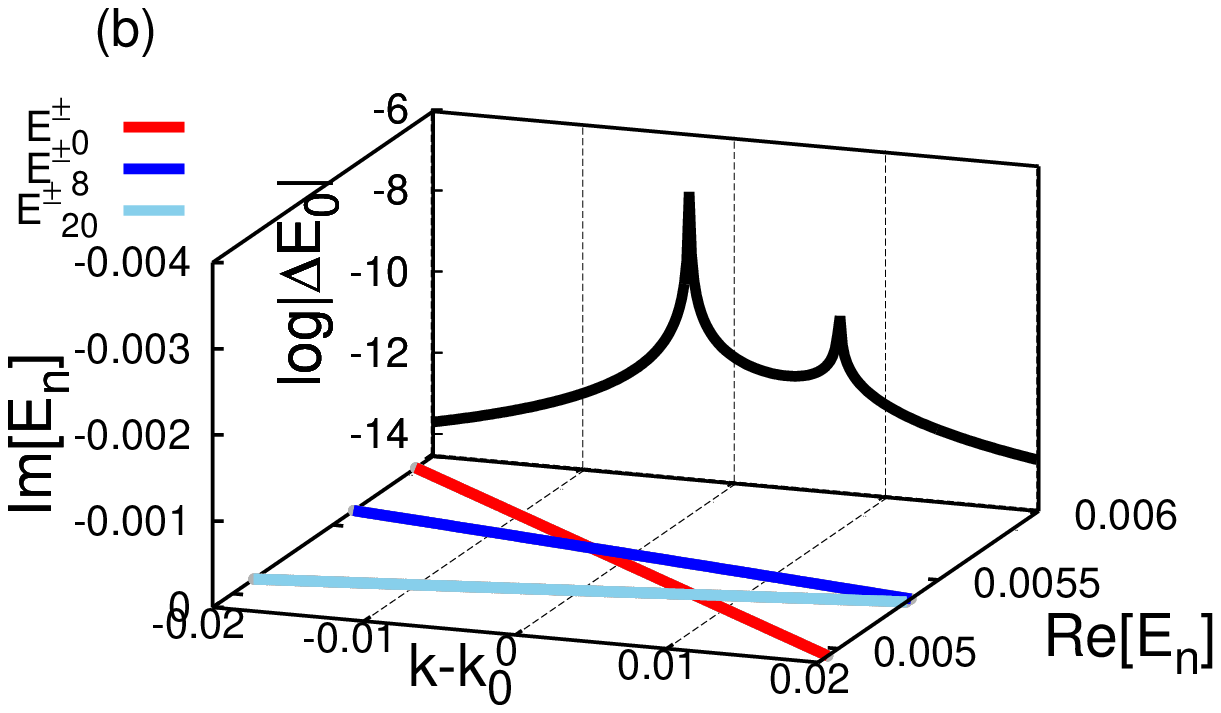}
	\includegraphics[width=0.4\textwidth]{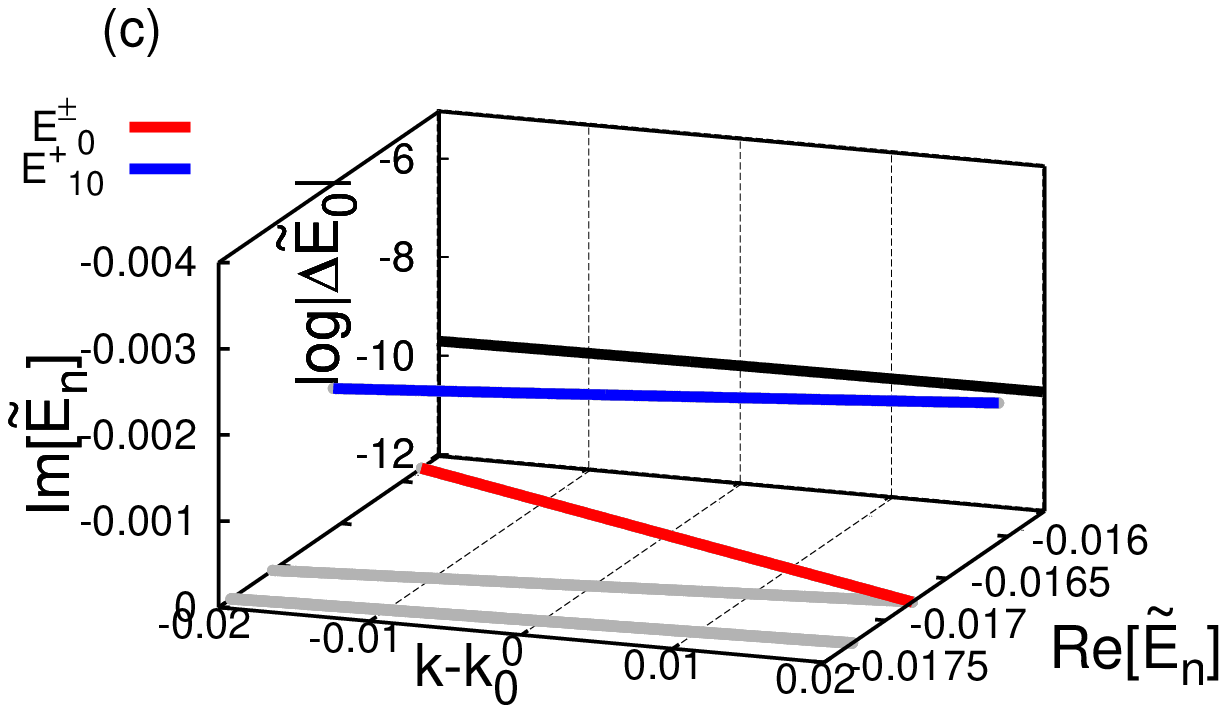}
	\includegraphics[width=0.4\textwidth]{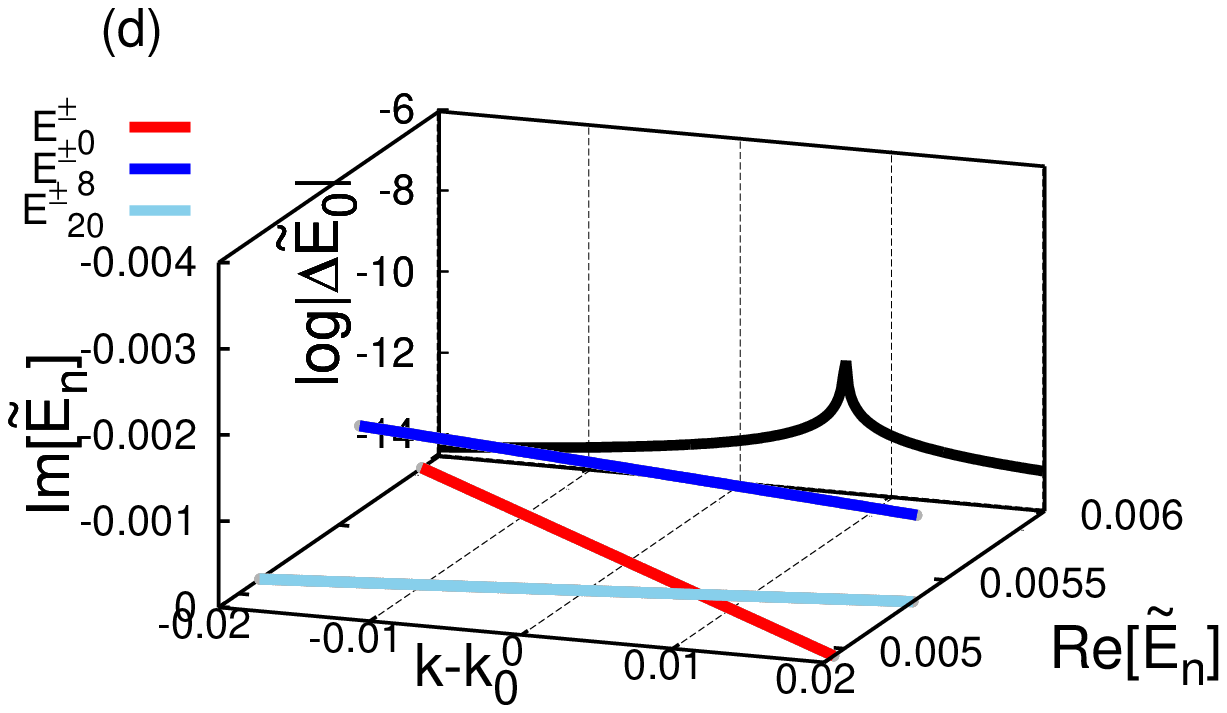}
	\caption{ \label{fig:spect}
	(Color online)
	Energy splittings $\Delta E_0$ (black lines in the back panel) and
	the reference doublet (red lines in the floor panel) and the related third state energies 
	(blue and light blue lines in the floor panel)
	as a function of the parameter $k$. 
	For each curve, the corresponding energy level is put in the figures.  
	Gray lines indicate energies of the states irrelevant to creating spikes. 
	In the upper panels, no absorber was applied for 
	(a) $1/h=27$ and (b) $1/h=44$. 
	In the lower panels, the strength of the absorber is set as $\Gamma=1$
	for  (c) $1/h=27$ and (d) $1/h=44$. 
	The index set $L$ is given as (a) $L = \{10\}$ and (b) $L = \{8\}$. 
}
\end{figure*}

Figures \ref{fig:spect}(a) and (b) demonstrate the splitting $\Delta E_0$ (in the back panel), 
together with the behavior of doublet and the third state 
energies (in the floor panel) as a function of the parameter $k$. 
When a spike appears in the $\Delta E_0$ vs $1/h$ plot, 
there always exist spikes in the plot of $\Delta E_0$ vs $k$ nearby. 
However, even if a spike appears in the $\Delta E_0$ vs $1/h$ plot, 
it does not necessarily mean that one exactly hits a spike in 
the plot of $\Delta E_0$ vs $k$.
These figures reveal that 
the third state for $1/h = 27$ (at the end of the first plateau) is the 10-th
excited state $\ket{\Psi^{+}_{10}}$ and this state is, 
as shown in Fig. \ref{fig:husimi} (b),  localized on the unstable 
fixed point $(q,p)= (0,0)$. 
On the other hand,  $1/h = 44$ (in the middle of the second plateau), 
there appear two spikes in the range under observation, one 
is the doublet composed of the 8-th excited states $\ket{\Psi^{\pm}_8}$
whose symmetric state is shown in Fig. {\ref{fig:husimi}} (c),
and the other is also given as a doublet of 
excited states,  $\ket{\Psi^{\pm}_{20}}$ whose symmetric state is shown in Fig. {\ref{fig:husimi}} (d).
As demonstrated in Figs. \ref{fig:husimi} (c) and (d), 
both doublets $\ket{\Psi^\pm_8}$ and $\ket{\Psi^\pm_{20}}$
are supported by elliptic inner and transversal outer KAM curves, respectively. 

\subsection{Absorbing operator}
\label{subsec:absorbing_operator}

Since the present situation, the case of $\tau=1$, is not far from the integrable limit, 
eigenstates $\ket{J_{\ell}}$ in the integrable limit well approximate 
eigenstates $\ket{\Psi_n}$, {\it i.e.}, $\bracket{J_\ell}{\Psi_n} \approx 1$. 
On the basis of this observation,  we introduce the following 
absorbing operator \cite{kaplan_1999,lippolis_2012}
\begin{equation}\label{eq:absorb_op}
\hat{P} = \hat 1-\frac{\Gamma}{2}\sum_{\ell \in L} \ketbra{J_\ell}{J_\ell}. 
\end{equation}
Here $\Gamma\le 2$ represents the absorbing strength, 
and $\hat 1$ the identity operator. 
The summation runs over a given index set $L$, 
which we choose appropriately depending on which states we want to suppress
\footnote{
In Eq. (\ref{eq:absorb_op}) we drop suffices $\pm$ since the definition of
$\hat P$ makes sense irrespective of the symmetry.}.
Below we consider the right eigenvalue problem 
for the absorbed (non-unitary) evolution operator
\begin{equation}\label{eq:non-unitary_eivenvalue}
\AbsU\ket{\tilde \Psi_n}=\tilde{u}_n\ket{\tilde \Psi_n}, 
\end{equation}
where
\begin{equation}
\AbsU = \hat{P}\hat{U},
\end{equation}
and $\tilde{u}_n = e^{-i\tilde{E}_n\tau/\hbar}$.
The following argument holds even if one 
considers the left eigenvalue problem.
% in the following discussion.

%Before numerically solving the eigenvalue equation (\ref{eq:non-unitary_eivenvalue}), 
First we will discuss what we can expect in perturbation theory with respect to 
the absorbing strength $\Gamma$. 
It is easy to show that a standard perturbative calculation up to the second order provides 
\begin{subequations}
\begin{equation}
  \tilde{u}_n \simeq u_n \cdot z_n,
\end{equation}
where
\begin{multline}\label{eq:perturbation_eval}
  z_n = 1 - \frac{\Gamma}{2}\sum_{\ell\in L} |a_{\ell,n}|^2 + \\
    \frac{\Gamma^2}{4}\sum_{\ell\in L} 
      \sum_{m\neq n}\frac{|a^\ast _{\ell,n} a^{}_{\ell,m}|^2}{u_n - u_m}u_m , 
\end{multline}
and 
\begin{equation}\label{eq:coefficient_1}
  a_{\ell,n} = \bracket{J_\ell}{\Psi_n}.
\end{equation}  
\end{subequations}
The right (absorbed) eigenstate is also given as 
\begin{subequations}
\begin{align}\label{eq:perturbation_evec}
  \ket{\tilde \Psi_n} &\simeq \ket{\Psi_n} - \frac{\Gamma}{2}
  \sum_{m \neq n} B_{m,n}\ket{\Psi_m}\\
  &=\ket{\Psi_n} - \frac{\Gamma}{2}\sum_{k=0}^{N-1}\sum_{m\neq n} B_{m,n} a_{k,m}\ket{J_k},
\end{align}
where
\begin{equation}
B_{m,n} = \sum_{\ell\in L} \frac{a^{\ast}_{\ell, m} a^{}_{\ell,n}}{u_n-u_m}u_m.
\end{equation}
\end{subequations}

For $0 <\Gamma\le2$, 
(absorbed) quasi-energies {$\tilde{E}_n$}
are no more real  because 
the second-order term in the perturbation expansion Eq. (\ref{eq:perturbation_eval}) becomes complex 
while the first-order term is real-valued. 
The eigenvalue $\tilde u_n$ is then shifted as 
\begin{equation}
\arg \tilde{u}_n=\arg u_n + \arg z_n. 
\end{equation}

By applying the absorber, the coupling between 
the absorbed and the rest of eigenstates is suppressed. 
This could be regarded as an inverse procedure of what 
is done in typical perturbation theory such as RAT theory, in which 
one starts with some unperturbed states $\ket{J_n}$ and build up
desired eigenstates $\ket{\Psi_n}$ by adding perturbation terms. 
The present absorbing method is, in a sense, to subtract perturbed terms $\ket{J_n}$
from the final state $\ket{\Psi_n}$. 
Therefore, applying the absorber in this way would be a test to check 
whether the final state $\ket{\Psi_n}$ could be obtained as a result 
perturbation in terms of unperturbed states $\ket{J_n}$, and, if so, 
which unperturbed states are involved in the perturbation procedure. 
The present absorbing method is equivalent to the one used in the open quantum systems,
e.g., \cite{keating_2008,laeck2010,backer_direct_2010,normann},
in which the absorbers are adopted as the Heaviside step function
$\bracket{x}{J}=H(x)$ or the Dirac delta function $\bracket{x}{J}=\delta(x)$.

The efficiency of the absorbing method is demonstrated 
in Figs. \ref{fig:spect}(c) and (d).
For absorbed quasi-energies, the tunnel splitting {$\Delta \tilde E^\pm_n$} is defined as 
{$\Delta \tilde E_n= \tilde E^-_n - \tilde E^+_n$},
however we note that {$\tilde E^\pm_n$} has an imaginary part when {$\Gamma>0$}. 
Each corresponds to the case where the absorber with $\Gamma=1$ is 
applied to the case shown in Fig. \ref{fig:spect}(c) and (d), respectively. 
Here the index set $L$ is chosen as $L=\{10\}$
in the case of {$1/h=27$}, 
and $L=\{8\}$ whose member corresponds to the doublet of 
the symmetric and  anti-symmetric state $\ket{J^\pm_8}$ for $1/h=44$. 
Here $\{\ell\}$ represents $\ket{J^{\pm}_\ell}$, and the member $\ket{J^\pm_\ell}$
in the index set $L$ is chosen in such a way that 
it maximally overlaps with the third state that is 
interacting the reference doublet $\ket{\Psi_0^{\pm}}$ and responsible for 
creating the spike. 

As clearly shown, energies of the associated third states gain certain 
amount of the imaginary part and 
pushed out to the complex plane, resulting in vanishing the spikes.
The effect to the other states is almost negligible.
However, as seen in Fig. \ref{fig:spect}(d), 
the right-hand peak with shorter height still remains since 
we have not include the states $\ket{J^{\pm}_{20}}$ in the absorber. 
As mentioned in the end of the previous subsection \ref{subsec:resonance}, 
there are two sets of doublets which are involved in avoided crossings in question.

\subsection{Staircase structure}
\label{subsec:tunnel_splitting}

In the previous subsection, we have selected out absorbing states 
by plotting energy levels around each avoided crossing and then judging 
by hand which states should be included in the set $L$, that is, 
it was necessary to refer to the figures like Fig. \ref{fig:spect}. 
We now introduce a systematic procedure to choose the absorbing states 
necessary to suppress the observed spikes. 

The most natural criterion to achieve this would be to check the energy difference 
from the reference doublet: 
$d(E^{\pm}_n) = | E^{\pm}_0 - E^{\pm}_n | \ (n=1,2,\cdots)$, 
because
the spikes appear when the reference doublet and a certain third state are energetically 
close to each other and form avoided crossings. 
We rearrange the states $\ket{\Psi^{\pm}_n}$
in ascending order of $d(E^{\pm}_n)$, and the corresponding  
integrable base $\ket{J^{\pm}_n}$ as well.  
The one-to-one correspondence between $\ket{\Psi^{\pm}_n}$ and $\ket{J^{\pm}_n}$ 
is again ensured since the condition 
$|\bracket{J^{\pm}_n}{\Psi^{\pm}_n}| \approx 1$
is now satisfied 
%not only %for $\tau=0.6$ but 
for $\tau=1$. 
Then the set of absorbing states containing the first $s$ doublets in the sense of 
the energy distance reads 
\begin{align}\label{eq:absorbing_set}
  L_s = \{  1, 2, \cdots, s \}, 
\end{align}
where we drop from the list of $L_s$ the states 
not belonging to the same parity as $\ket{\Psi_0^{\pm}}$.
We must recall that the ground state {$\ket{\Psi_0^{\pm}}$} 
has the symmetry with respect to the translation in addition to the mirror transformation.

Note that the set $L_s$ of absorbing states depends on the value of $h$, so 
has to be determined for each $h$.
As explained below, the reason why we consider the cases $s>1$
is that the energetically nearest state $L_1$ from the reference doublet
is not sometimes sufficient for killing the coupling with the reference doublet. 

\begin{figure}[t]
	\center
	\includegraphics[width=0.5\textwidth]{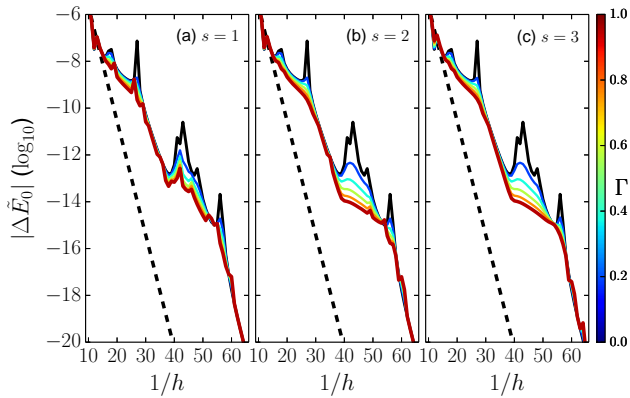}
	\caption{\label{fig:nonunitary_splitting}
	(Color online)
	Tunneling splittings $\Delta \tilde{E}_0$ for 
	the evolution operator $\AbsU$ are plotted as a function of $1/h$. 
	The range of interaction in increased as 
	(a) $s=1$, (b) $s=2$, and  (c) $s=3$.
	Black curves show the unitary case ($\Gamma = 0$) and 
	each colored curve the case with absorber, and the colors distinguish 
	the absorbing strength $\Gamma$ (see the right-hand color bar). 
	The splitting $\Delta E_0$ in the integrable limit 
	is shown as black dashed lines in each figure. 
}
\end{figure}

Figure \ref{fig:nonunitary_splitting} plots the splitting 
of quasi-energy $\tilde{E}^{\pm}_0$ evaluated 
for the operator $\AbsU$ as a function of $1/h$. 
In the $s=1$ case, we see that 
some spikes, especially in the first plateau, disappear with increase in $\Gamma$. 
As shown in Fig. \ref{fig:spect}, the absorber pushes the third level into the 
complex domain, and the coupling with the reference doublet is suppressed.

However, we notice that some spikes still remain in the second plateau regime. 
This is due to the fact that, as shown in Fig. \ref{fig:spect}(b), 
some spikes come close to each other 
in the second plateau and a single absorber is not enough 
to suppress the interaction with the reference doublet. 
In the case presented in Fig. \ref{fig:spect}(b), the third state responsible 
for the left-hand peak is the state supported by an elliptic torus inside the 
KAM region and the right-hand one is supported by a transversal torus. 
As we further add the corresponding absorbers in this way, 
the peaks surviving in the $s=2$ case gradually disappear, 
and the curve almost converges at $s=2$. 

%%% ShShSh(2014.10.7) 
It would be worth emphasizing that steeply decaying regions are not affected
and robust against the the absorber applied on plateaus.  
This strongly suggests that the influence of spikes is well localized in each plateau, 
not like the situation suggested in \cite{laeck2010}.
This observation also supports our hypothesis; the splitting curve should be viewed as 
a staircase structure accompanied by spikes, not as spikes bringing the staircase.

\section{Mechanism generating the staircase structure}
\label{sec:eigenstate}

%%% ShShSh(2014.9.23)  
The main message in the previous section is that 
the staircase-shaped skeleton is formed in the splitting curves %shown in Fig. \ref{fig:splitting}, 
and spikes are superposed on it. 
In this sense we may say that the origin of the enhancement of the tunneling probability 
traces back to such a staircase structure. 
In this section, we study the mechanism creating the staircase structure 
by introducing the renormalized basis, and show the reason why this 
only appears in nonintegrable systems.

\subsection{Instanton-noninstanton transition}\label{subsec:I-NI}

\begin{figure}[t]
	\center
	\includegraphics[width=0.45\textwidth]{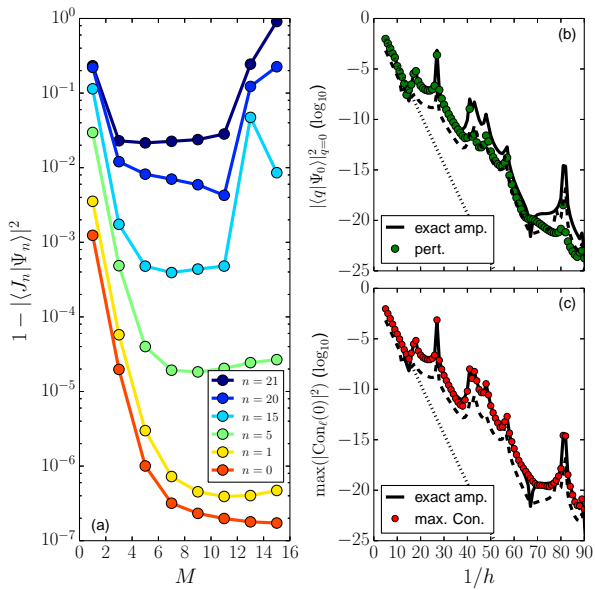}
	\caption{\label{fig:overlap}
	(Color online)
	(a) Left panel shows the error $1 - |\bracket{J^{(M)}_n}{\Psi_n}|^2$ 
	as a function of the BCH order $M$ in the case of $h=1/63$.  
	Not only the grand state, $n=0$, but also excited states up to $n=74$ are 
	examined. 
	(b1) The amplitude at $q=0$ of the ground state 
	obtained by the perturbation calculation, and 
	(b2) the amplitude of the maximal mode of the contribution spectrum (see the text). 
	The exact amplitude for the ground state $\ket{\Psi_0}$ at $q=0$ is shown 
	as blue curves in (b1) and (b2). 
	The solid black dashed and dotted curves show the
	exact level splitting $\Delta E_0$ and the level splitting $\Delta E_0^{(M)}$ of the integrable basis
	%%%%%%%%
	for reference, respectively.
}
\end{figure}

%%% ShShSh(2014.9.23) 
As shown in \cite{ikeda_2013}, the I-NI transition 
%in the tunneling process 
could be well captured by renormalized perturbation theory. 
An important finding there was that a remarkable quenching of renormalized transition
matrix elements explains the I-NI transition.
In particular, without using highly renormalized integrable Hamiltonian
as unperturbative bases one could not  identify the mechanism behind the transition. 

For this reason, we also apply the same perturbation scheme to pursue the origin of 
the staircase structure. 
In essence, renormalized perturbation theory makes use of 
the Baker-Campbell-Hausdorff (BCH) expansion \cite{scharf_1988,ikeda_2013}:
%%% EndEndEnd %%%
\begin{equation}
  \hat{U}  \approx 
  \hat{U}_M\equiv\exp\BRA{-\frac{i}{\hbar}\tau \hat{H}_\mathrm{eff}^{(M)}(\hat{q},\hat{p})},
\end{equation}
where  
\begin{equation}\label{eq:bch_hamiltonian}
  \hat{H}_\mathrm{eff}^{(M)}(\hat{q},\hat{p})
  =\hat{H}_1(\hat{q},\hat{p})+\sum_{\underset{(j\in \text{odd int.})}{j=3}}^{M}
  \Bra{\frac{i\tau}{\hbar}}^{j-1}\hat{H}_{j}(\hat{q},\hat{p}).
\end{equation}
Here $\hat{H}_j$ denotes the $j$-th order term in the BCH series. 
Explicit forms for the
first few terms are found as
\begin{subequations}
  \begin{align}
    \hat{H}_1(\hat{q},\hat{p}) &= T(\hat{q}) + V(\hat{p}),\\
    \hat{H}_3(\hat{q},\hat{p}) &=  \frac{1}{24}\big([T, [T, V]]-[V,[V, T]] \big),\\
    &\vdots \notag
  \end{align}
\end{subequations}
where the terms {$\hat{H}_j$} for even $j$ are equal to zero 
thanks to the symmetrized form of $\hat U$. 
The first order BCH Hamiltonian $\Heff^{(1)}$ is identical 
to the continuous time Hamiltonian and 
higher order BCH Hamiltonians $\BCH$ are expressed as nested commutators. 
We denote the eigenfunctions of the integrable Hamiltonian 
$\hat{H}_\mathrm{eff}^{(M)}$ by $\ket{J^{(M)}_\ell}$:
\begin{equation}
  \hat{H}_\mathrm{eff}^{(M)}\ket{J^{(M)}_\ell}
  =  E_{\ell}^{(M)} \ket{J^{(M)}_\ell}.
\end{equation}

\begin{figure}[b]
	\center \includegraphics[width=0.45\textwidth]{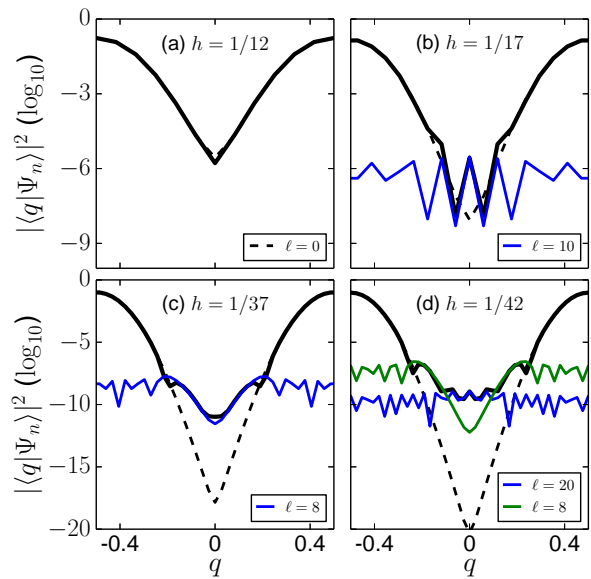}
	\caption{\label{fig:eigen}
	(Color online)
	The black curve in each figure shows the eigenstate $\ket{\Psi_0}$
	for $\tau=1$	in the (a) instanton, (b) first plateau, 
	(c) second decay and (d) second plateau regime, respectively.
	The dashed curve displays the integrable eigenstate 
	$\bracket{q}{J^{(M)}_0}$ at the corresponding $h$ value, 
	and colored ones the integrable components 
	$\bracket{q}{J^{(M)}_\ell}\!\bracket{J^{(M)}_\ell}{\Psi_0}$ at $q=0$, 
	where the value of $\ell$ is put in each figure. Note that the structure around $q=0$ can be well 
	reproduced by the maximal mode(s) of the contribution spectrum (see the text). 
}
\end{figure}

We first check the validity and efficiency of renormalized perturbation bases by 
examining the error $1 - |\bracket{J^{(M)}_n}{\Psi_n}|^2$ of the approximation.
As shown in Fig. \ref{fig:overlap}(a), 
the BCH states becomes better approximation to the corresponding 
eigenstate $\ket{\Psi_n}$ as the expansion order $M$ increases.
Note also that the expansion works for the lower energy eigenstates as compared to 
the higher excited states. 
This is, however, not a convergent expansion: the error $1 - |\bracket{J^{(M)}_n}{\Psi_n}|^2$ 
starts to grow when the expansion order $M$ exceeds a certain optimal order. 

Such highly efficient integrable approximation ensures the validity of renormalized perturbation, 
in which the difference 
$\Delta \hat{U}_M = \hat{U} - \hat{U}_M$ could be regarded as a perturbation \cite{ikeda_2013}.
As also shown in Fig. \ref{fig:overlap}(b1), the results of the 1st order perturbation calculation 
are in an excellent 
agreement with the exact ones, and even the staircase structure could be 
reproduced. However we would like to remark that although perturbation theory, 
not necessarily the present one, works well, 
this does not tell us anything about the underlying mechanism generating the staircase.

As shown in Fig. \ref{fig:overlap}(b1), the splitting $\Delta E$ is strongly correlated with the amplitude 
of the eigenstate at $q=0$, and characteristic patterns appear around $q=0$.
As seen in Fig. \ref{fig:eigen}(a), the eigenstate $\ket{\Psi_0}$ for $\tau=1$ 
in the instanton regime is,
as expected, well fitted by the one $\ket{J_0^{(M)}}$ in the integrable bases, 
whereas the integrable approximation does not work any more and 
further structures appear in other regions.
In the first and second plateau, 
%there appears the bulge at $q=0$ and 
the curve bends in a convex way
(see Fig. \ref{fig:eigen}(b) and (d)),
but in the first steeply decaying region the curve bends in a downward direction
at $q=0$ and takes a concave structure
(see Fig. \ref{fig:eigen}(c)).

To explore the nature of wavefunctions at $q=0$, 
we here introduce a spectrum decomposition at each position $q$ 
in terms of integrable bases $\ket{J^{(M)}_\ell}$: 
\begin{equation} \label{eq:coeff_expand}
  \bracket{q}{\Psi^{+}_0} = \sum_{\ell=0}^{N-1} \Con(q)
\end{equation}
where
\begin{equation} \label{eq:cont}
  \Con(q)= \bracket{q}{J^{(M)}_\ell}\!\bracket{J^{(M)}_\ell}{\Psi_0^{+}}.
\end{equation}
We call such a decomposition the {\it contribution spectrum} \cite{ikeda_2013}.
In the following discussion, we focus only on the symmetric ground state $\ket{\Psi_0^{+}}$.
As we mentioned in Sec. \ref{sec:model}, 
each eigenstate has a symmetry with respect to the mirror transformation $\hat{\Pi}_q$ and 
the translation $\hat{T}_r$.
Therefore, the basis $\ket{J^{(M)}_\ell}$ that has the same symmetry as $\ket{\Psi_0^{+}}$
is only used for the contribution spectrum.
%and the zero-th component of $\bracket{J^{(M)}_\ell}{\Psi_0^+}$
%belonging to different symmetry classes will be ignored. 

\begin{figure}
	\centering
	\includegraphics[width=0.40\textwidth]{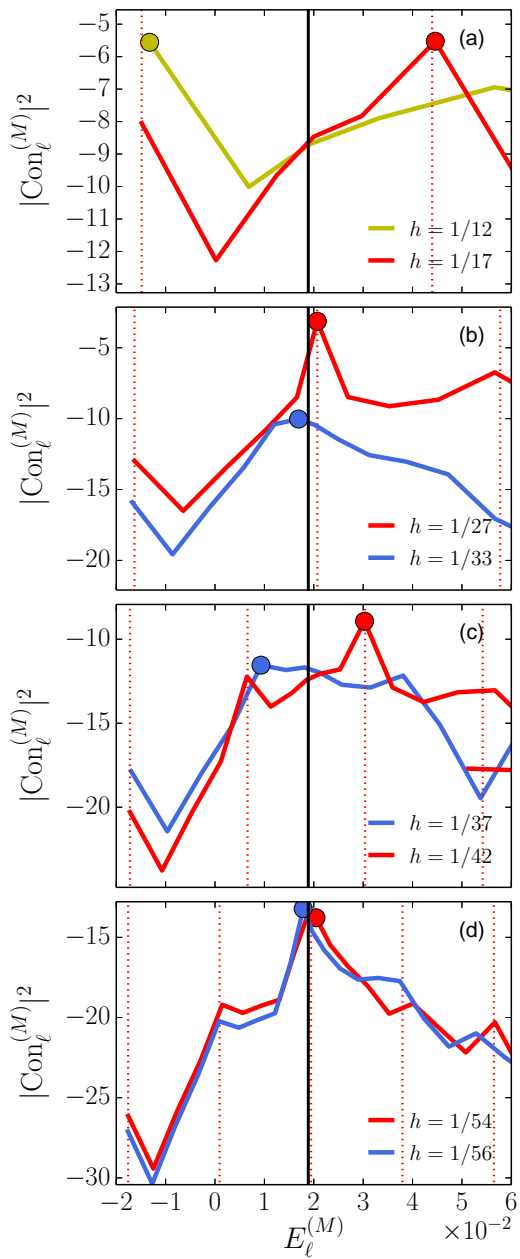}
	\caption{\label{fig:cont}
	(Color online)
	The contribution spectrum $\Con$ (in $\log_{10}$ scale) at $q=0$ 
	is plotted as a function of the energy $E_{\ell}^{(M)}$. 
	The BCH order $M=7$ was used. 
	The values of $h$ are indicated in each figure.  
	Each panel respectively shows the case 
	(a) before (yellow) and after (red) I-NI transition, 
	(b) before (red) and after (blue) the transition from the first plateau to the second decaying, 
	(c) before (blue) and after (red) the transition from the second decaying to the second plateau,
	(d) before (red)  and after (blue) the transition from the second plateau to third steep decaying regime.
	The dot represents the maximal mode in each spectrum. 
	We have used a yellow-colored curve in the region where the maximal mode is given by 
	the instantion contribution, a red-colored when the maximal mode energy is above the 
	separatrix energy, and a blue-colored below the separatrix energy. 
	The thick solid line represents the separatrix energy and red dotted lines the energies 
	satisfying the condition $E = E^{(M)}_0 + mh/\tau$ ~~ $(m=0, 1,2, \cdots)$
	(see the discussion in subsection \ref{subsec:action}). 
}
\end{figure}

As shown in Fig. \ref{fig:overlap}(b2), since the maximal mode of the contribution spectrum 
quite efficiently describes the behavior of the splitting, % as a function of $1/h$, 
we can deduce that the staircase structure must be characterized 
by the maximal mode. 
Indeed, in Ref. \cite{ikeda_2013}, we have shown that the instanton-noninstanton (I-NI) transition 
could be explained as the switching behavior of 
the most dominant component in the contribution spectrum;
from the one representing the instanton contribution to broad 
components supported around the separatrix of a central unstable fixed point. 
%This was first discovered in the H\'enon map \cite{ikeda_2013}. 
Below, we present that the dominant component controls 
not only the transition from instanton to noninstanton but overall signatures  
in the staircase structure. 
We will explain this by showing contribution spectra for several values of $1/h$, which are 
presented in Fig. \ref{fig:cont}.

First of all, as mentioned just above, 
we notice that the contribution spectrum is mainly composed of two peaks 
with distinct characteristics.  The first one is a sharp peak located at $E = E_0^{(M)}$, 
and the second is composed of many components, whose center is 
situated around the separatrix energy.
A small peak sometimes appears on the broadly 
spread components as a result of the interaction with a third state. 
The first sharp peak at $E = E_0^{(M)}$ originates from the instantion contribution 
that has a maximal overlap with the ground state $\ket{\Psi^+_0}$, 
so we hereafter call it the instanton peak.
We stress again that the instanton peak at $E = E_0^{(M)}$ can be recognized only when 
we prepare higher order BCH expansions ($M=7$ for the present calculation), 
otherwise the instanton peak is not isolated from the others and
could not be identified. 

In the instanton decay (the first steeply decaying) regime, which is
seen in the case of $1/h$  = 12 in Fig. \ref{fig:cont}(a), 
the instanton peak dominates the other components. 
As a result the amplitude of the ground state $\ket{\Psi^+_0}$ at $q=0$ is well described by 
the integrable Hamiltonian base $\ket{J^{+, (M)}_0}$.
Hence the instanton behavior should and is actually observed.

With increase in $1/h$, the height of both peaks, 
the instanton peak and the broad components centered around the separatrix, 
gradually drop, but the speed of the former is much higher than 
that of the latter, eventually resulting in the switching of the role of the dominant contributor 
from the instanton to the top of broad components. 
An important remark is that the support of the state associated with 
the top of the broad components is outside the separatrix, meaning that 
the ground state is most dominantly coupled with an outside state \cite{ikeda_2013}. 
We notice in Fig. \ref{fig:overlap}(b) that, exactly at this switching moment, the first instanton 
decay turns to the first plateau,
and eigenstates show the convex structure around $q=0$ (see Fig. \ref{fig:eigen}(b)). 
In the perturbation calculation, 
it is also crucial to include outer torus states into unperturbed bases 
to reproduce the convex structure at $q=0$, 
otherwise the resulting wavefunction cannot bend upward at 
$q=0$ since it is merely a superposition of exponentially decaying states.  
It is important to note that not only 
such a convex structure just after the transition but also neighboring structures
around $q=0$ could be well reproduced only by the maximal mode in the spectrum $\Con$
(see Fig. \ref{fig:overlap}(b2)). 

\begin{figure}[t]
	\centering
	\includegraphics[width=0.4\textwidth]{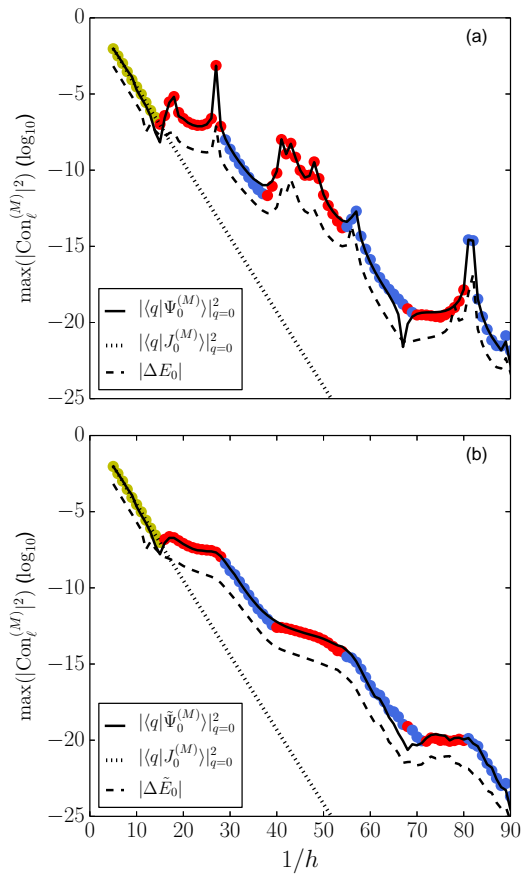}
	\caption{ \label{fig:contmax_and_splitting}
	(Color online)
	(a) Maximal modes plotted as a function of the inverse Planck's constant $1/h$.
	We have used yellow-colored dots in the region where the maximal mode is given by
	the instanton contribution, red-colored ones when the maximal mode energy is above the
	separatrix energy, and blue-colored below the separatrix energy.
	(b) Maximal modes for the absorbed eigenstates $\ket{\tilde \Psi_n}$.
	The absorption procedure is the same as that introduced in section \ref{sec:staircase}. 
	Here we use the integrable bases $\ket{J_\ell^{(M)}}$ as the absorber, 
	and the absorption parameters are chosen as $s=3$ and $\Gamma=0.4$.
	The rule for color coding is the same as in (a).
	In both calculations, the BCH order $M=7$ was used.
	In (a) the exact eigenfunction $\bracket{q}{\Psi_0^+}$ at $q=0$, 
	integrable basis  $\bracket{q}{J_n^{(M)}}$, and energy splitting $\Delta E_0$ are shown as solid,
	dotted and broken curves, respectively. 
	In (b) the solid curve represents the absorbed eigenfunction 
	$\bracket{q}{\tilde \Psi_0^+}$ at $q=0$, 
	and dotted and broken ones are the same as in (a).
}
\end{figure}

In any case, the maximal mode in the spectrum $\Con$ can be a good indicator 
for the value of eigenstates at $q=0$ and thereby the splitting $\Delta E$. 
%As shown in Fig. \ref{fig:contmax_and_splitting}(a), 
%this holds not only for the grand state but also for exited states. 
The maximal mode in the contribution spectrum $\Con$, which is shown using 
color-coded dots in the Fig. \ref{fig:contmax_and_splitting},  well traces the staircase 
structure of the exact splitting $\Delta E_n$, and the value of eigenstates 
$|\bracket{q}{\Psi_n}|^2_{q=0}$ at $q=0$ as well. 
We will fully make use of this fact hereafter.

As we further increase in $1/h$, the instanton peak is completely overtaken by the 
broadly spread  components  (see Fig. \ref{fig:cont}(b)) and this ordering is fixed and never turned over. 
%%%ikikikM>3
We also emphasize that the estimation of the critical Planck's constant $h_c$ at which the I-NI transition 
occurs becomes a bit imprecise if we use the lower order BCH series. 
%In the current situation we need the expansion at least with $M > 3$. 
%%%%%%%%%%%

As we increase $1/h$ after the I-NI transition, 
the support for the maximal mode of the contribution spectrum further approaches the separatrix, 
which is shown as in Fig. \ref{fig:cont}(b), and 
eventually it goes into the inner tours region in excess of the separatrix. 
At this moment, we realize that the splitting curve changes the behavior
from the first plateau to the second steeply decaying regime (see Fig. \ref{fig:overlap}(b)). 
At the same time, the structure of eigenstates at $q=0$ changes 
from the convex to concave shape (see Fig. \ref{fig:eigen}(c)). 
%
%of the contribution spectrum monotonically approaches the separatrix in the fist plateau regime, eventually it goes into the inner tours region in excess of the separatrix.  We also realized, at this moment, the splitting changes the behavior from first plateau to second steep decay regime (see Fig. \ref{fig:cont}(b)) and the local structure of eigenstates at $q=0$ changes the concave structure.

With further increase in $1/h$
the maximal mode also shifts to the left. 
On the other hand, another peak is born at the right-hand edge of broad components, 
and now the competition comes into issue between the those peaks, the one playing a major role in 
the I-NI transition, and the new one at the right-hand edge. 
As noticed in Fig. \ref{fig:cont}(c), the switching of the dominant contributor 
again takes place between these two peaks, and at this moment the splitting curve 
turns from the second steeply decaying to the the second plateau regime.

After such a transition, the overtaken peak, the one playing a role in 
the I-NI transition, is gradually absorbed into the spectrum envelope.
However it leaves a clear trace in wavefunction: 
As shown in Fig. \ref{fig:eigen}(d), the shoulder or bulge observed in the neighboring 
region around $q=0$ is well reproduced by the component that has played a role in 
the I-NI transition. The convex structure observed in the first plateau 
is pushed outward by the newly born component, and then it appears as shoulders. 
In other words, the history of the staircase structure in the splitting plot is properly 
recoded in the tail of wavefunction, not necessarily at $q=0$.

The staircase structure in the splitting plot could therefore be explained 
by the successive switching process of maximal modes, and passing through the 
separatrix, that is, whether
%%%ikikikseparatrixseparatrix%%%
the support of the maximal mode is inside or outside the separatrix. 
%%%%%%
%%%ikikik
%Figure \ref{fig:contmax_and_splitting} displays the splitting curve, for which 
%we put two colors depending on which peak is dominant and whether the peak 
%is inside or outside the separatrix.  
Figure \ref{fig:contmax_and_splitting} illustrates that the staircase structure 
of the splitting curve can be understood by the position of the maximal mode: 
whether its support is outside or inside the separatrix.

We have verified, as shown in Fig. \ref{fig:contmax_and_splitting}(b), that even if we suppress the peak 
standing on the broad peak components using the same absorber technique 
in subsection \ref{subsec:absorbing_operator}, this switching process still survive. 
This implies that the switching does not occur specifically between the resonance peaks 
appearing in the contribution spectrum, but overall deformation of broad peak components
controls it.
We could identify at least the third and fourth transition and confirmed the same scenario applies.

\subsection{Anomaly of eigenfunctions in the action representation}\label{subsec:action}

As shown above, we could attribute the emergence of the staircase structure 
to the successive switching of the dominant component in the contribution spectrum. 
%%%ikikik
%In this subsection, we explain why the dominant component gradually shifts with 
%increase in $1/h$ and 
%also the reason why the slope of the splitting curve  
%changes at each moment  when the dominant modes passes through the separatrix. 
In this subsection, we explain why the quantum number of the dominant component 
gradually shifts with increase in $1/h$, passing through the separatrix,
%%% ShShSh(2014.11.4) %%% 
%state, 
and also explain why 
%to pass through the separatrix
this causes the change in the slope of the splitting curve.

For this purpose, we examine the behavior of the expansion coefficient 
$\bracket{J_\ell^{(M)}}{\Psi_0}$ and the integrable eigenfunction $\bracket{q}{J_\ell^{(M)}}$ at $q=0$ separately. 
Note that the product of these two terms constitutes each component in the contribution spectrum $\Con$. 
We here call $\bracket{J_\ell^{(M)}}{\Psi_0}$ the eigenfunction in the action representation. 
%since the bases $\{\ket{J_\ell^{(M)}}\}$ are an integrable approximation 
%to the quantum map (\ref{eq:qmap}).
%

\begin{figure}[t]
	\centering
	\includegraphics[width=0.45\textwidth]{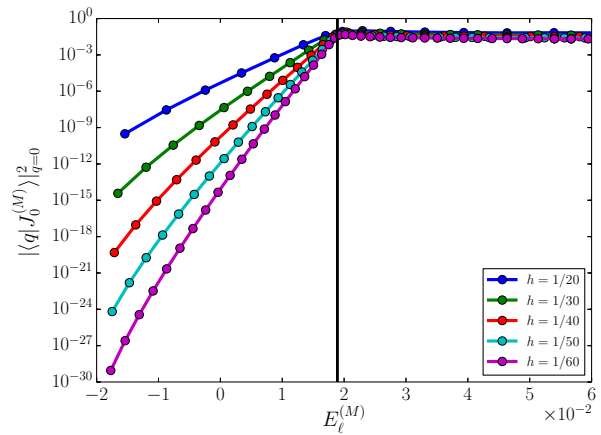}
	\caption{\label{fig:irep}
	(Color online)
	Amplitude of the integrable eigenfunction $\bracket{q}{J_\ell^{(M)}}$ at $q=0$. 
	The 7th order BCH Hamiltonian was used. 
	The black vertical line shows the energy of separatrix. 
}
\end{figure}

First of all we remark that the value $h/\tau$ becomes a fundamental energy unit in our system. 
This is because the present system is driven by the periodic force with period 
$\tau$, so $2\pi/\tau \times \hbar = h/\tau$ becomes a fundamental energy unit, 
and the energies specified 
as $E_\ell^{(M)} = E_0^{(M)} + m h/\tau ~~(m = 0,1,2,\cdots)$ may 
invoke quantum mechanical resonances. In Fig. \ref{fig:cont}, we have shown such 
energies as dotted red lines. 
%%%%% EndEndEnd %%%%%%
In the following, we first describe a signature of $\bracket{q}{J_\ell^{(M)}}$ at $q=0$ and 
then discuss anomaly found in $\bracket{J_\ell^{(M}}{\Psi_0}$.
Combining these, we finally explain the mechanism of successive switching in the contribution spectrum. 
%As noticed below, the energies specified above indeed play a key role. 

%
%separated from
%the energy $E_0^{(M)}$ by the integer multiple of the 
%fundamental energy unit $h/\tau$ mentioned above,
%%%%
%which plays a crucial role as discussed below. 
%%ikikik
%In Fig. \ref{fig:irep}, the thick black line indicates the energy of separatrix. %position of the separatrix. 

%%%%%%%%%%%%%%%%%%%%%%%%%%%%%%%%%%%%%%%$$$$$$$$$$$$$$$$

As shown in Fig. \ref{fig:irep}, 
the amplitude of the integrable eigenfunction $\bracket{q}{J_\ell^{(M)}}$ at $q=0$ 
shows exponential dependence on %decays exponentially as a function of 
the energy $E_\ell^{(M)}$ as far as the energy is less than that of 
the separatrix (left side of the thick black line in Fig. \ref{fig:irep}). 
When plotting $\bracket{q}{J_\ell^{(M)}}$ with a fixed energy one also 
finds exponential decay as a function of $1/h$ (see Fig. \ref{fig:contmax_and_splitting}(b)). 
This is an expected behavior since $\bracket{q}{J_\ell^{(M)}}$ is just an eigenfunction of 
an integrable Hamiltonian, no matter large the expansion order $M$ is.

%On the other hand, above the separatrix energy (right side of the thick black line), 
%we see that the integrable eigenfunction $\bracket{q}{J_\ell^{(M)}}$ keeps constant. 
%This is also reasonable because $\bracket{q}{J_\ell^{(M)}}$ has its supports on 
%the transversal invariant torus, which contains the point $q=0$. Hence the point $q=0$ is 
%not in the tunneling tail, but in the region having the real support, resulting in 
%no decay. 

On the other hand, above the separatrix energy (right side of the thick black line), 
we see that the amplitude of the integrable eigenfunction $\bracket{q}{J_\ell^{(M)}}$ 
keeps almost constant. 
This is also reasonable because each $\bracket{q}{J_\ell^{(M)}}$ has its supports on
a transversal invariant torus outside the separatrix, so the connection 
is not made via tunneling but real classical processes, 
thus resulting in no decay as a function of the energy.

\begin{figure}[t]
	\centering
	\includegraphics[width=0.42\textwidth]{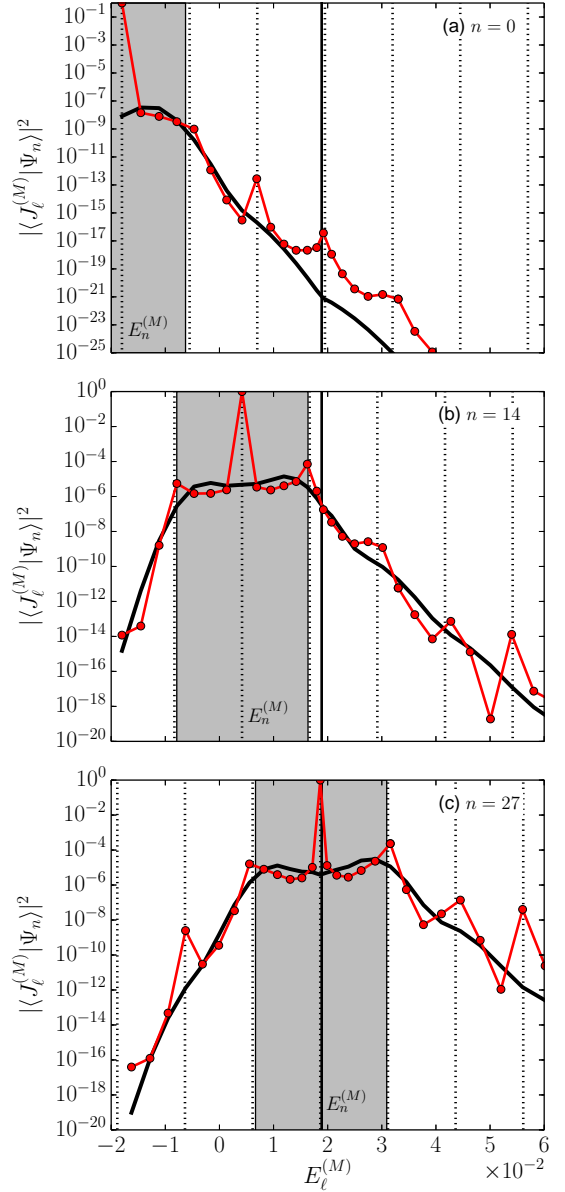}
	\caption{\label{fig:action_rep}
	(Color online)
	Eigenstates $\bracket{J_\ell^{(M)}}{\Psi_n}$ in the action representation 
	plotted as a function of $E^{(M)}_\ell$
	for $h=1/80$ for 
	(a) $n=10$, (b) $n=15$ and (c) $n=25$, respectively.
	%Here we use the 7-th order BCH Hamiltonian for the basis $\ket{J_\ell^{(M)}}$.
	The black curves show the matrix elements $\bra{J_\ell^{(M)}}\Delta \hat{U}\ket{J_n^{(M)}}$.
	Here we used the 7-th order BCH Hamiltonian as the basis $\ket{J_\ell^{(M)}}$.
	The black solid line and dotted lines respectively show the separatrix energy, and 
	the energies satisfying the condition 
	$E = E_0^{(M)} + m h/\tau ~~(m = 0,1,2,\cdots)$. 
}
\end{figure}

In contrast, the nature of the eigenfunction 
$\bracket{J_\ell^{(M)}}{\Psi_0}$ in the action representation is highly nontrivial.
As shown in Fig. \ref{fig:action_rep}(a), 
there exists a sharp peak at $E_0^{(M)}$, 
which represents the instanton contribution, 
and then the value of $\bracket{J_\ell^{(M)}}{\Psi_0}$ suddenly drops 
to reach a small level. 
Then it forms a {\it non-decaying region}
in which %the contribution 
the value of $\bracket{J_\ell^{(M)}}{\Psi_0}$ does not decrease, 
rather increases gradually until a small  
peak which is close to the energy which is specified 
by the relation  $E=E_0^{(M)} + h/\tau$ \cite{ikeda_2014,hanada_2015}. 
This peak originates from the resonance of the associated states with the periodic forcing 
inherent in our model. The non-decaying region means 
that as long as the eigenphase difference is less than $h/\tau$
the contribution from the associated integrable basis states is almost equal.
It is beyond this resonance that the exponential decay common in the ordinary tunneling 
tail takes place. 
%In this sense, we have to say that 
%the non-decaying behavior inside the resonance is anomalous. 
As presented in Figs. \ref{fig:action_rep}(b)-(c), %such anomaly 
the presence of non-decaying region	
is not limited to the ground state but appears in exited states as well.  
Also note that overall features are reproduced by just one-step time evolved 
wavefunciton which is expressed as $\bra{J_\ell^{(M)}}\Delta U_M\ket{J_n^{(M)}}$. 
The latter is consistent with the observation that perturbation theory based on 
the BCH basis works well (see Fig. \ref{fig:overlap}(b)). 
We emphasize that these are all observed only when the order $M$ 
of the BCH approximation is large enough 
and also universally appear in the eigenfunction 
of quantum maps \cite{ikeda_2013,ikeda_2014,hanada_2015}.  

\begin{figure}[t]
\centering
\includegraphics[width=0.45\textwidth]{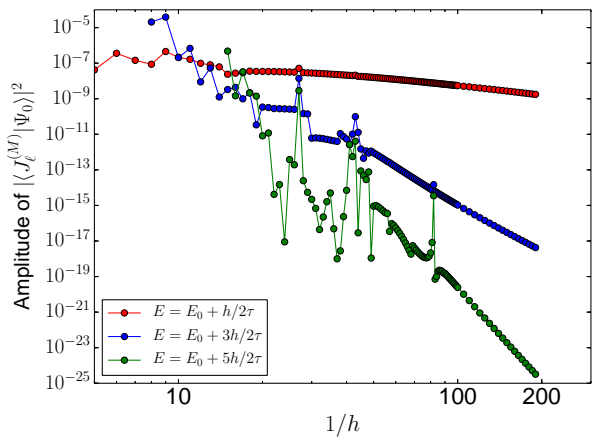}
\caption{\label{fig:irep_amp}
	(Color online)
	The inverse Planck's constant $1/h$ dependence of $\bracket{J_\ell^{(M)}}{\Psi_0}$. 
	Difference of colors distinguishes the energy at which the value of 
	$\bracket{J_\ell^{(M)}}{\Psi_0}$ is evaluated (see the inner panel). 
}
\end{figure}

A particularly important fact is, as shown in Fig \ref{fig:irep_amp}, that 
the decay rate of the height of the non-decaying region 
as a function of $1/h$ is extremely slow, as compared to 
the region $E_\ell^{(M)} > h/\tau$. 
This clearly distinguishes and characterizes the two regions, 
below and above the resonance energy $E=E_0^{(M)} + h/\tau$. 
%and also explains the generation of the staircase structure as stated below. 
We should make clear the underlying reason behind the observed power law decay 
in both regions, but the observed energy, so the corresponding classical structure as well, 
moves with increase in $1/h$ in the current setting, which makes 
difficult to apply a straightforward semiclassical argument. 
%However, it is highly possible that these are the 
%phenomena beyond the leading order semiclassical description 
%because of its decaying nature \cite{hanada_2015}. 

In addition to the resonance peak at $E = E_0^{(M)} + h/\tau$, 
a sequence of peaks implying the higher order resonances appear 
at $E = E_0^{(M)} + mh/\tau$ ($m$ integer) (see Fig. \ref{fig:action_rep}(a)). 
In conjunction with resonance peaks,
there also exist narrow non-decaying regions just below each peak 
as the non-decaying region appearing in the region $E - E_0^{(M)} < h/\tau$.  
Such a sequence of non-decaying region is not so sharply identified 
in Fig. \ref{fig:action_rep}(a), but it becomes clearly visible as we increase $1/h$. 
We can therefore divide each sector $E_0^{(M)} + m h/\tau < E < E_0^{(M)} + (m+1)h/\tau$
into two characteristic regions; the one showing faster decay with $1/h$ and the other having 
quite slow decaying character.  
A detailed explanation will be presented in our forthcoming paper \cite{hanada_2015}, 
and we just show in Fig. \ref{fig:irep_amp} the difference of the decay rate 
by measuring it in the middle energy in each sector. As is seen, 
the decay rate in the region $E_0^{(M)} + h/\tau < E < E_0^{(M)} + 2h/\tau$ 
is much slower than in the next sector $E_0^{(M)} + 2h/\tau < E < E_0^{(M)} + 3h/\tau$. 
Although we do not specify in which characteristic region the middle point energies 
used to measure the decay rate is contained, it is enough, 
in the following argument, to notice that the decay rate much differs in each sector. 
Also note that such resonance peaks with the same nature 
also appear in exited states as also shown in Fig. \ref{fig:action_rep}(b)-(c). 

Putting all the pieces together, we can now understand why successive switching 
in the contribution spectrum generates the staircase structure. 
In the first decaying (instanton) region, 
the instanton is the most dominant
and broadly spread components provide only negligible contributions, as explained 
in the previous subsection. 
The height of the instanton peak decays exponentially 
with $1/h$ as expected. 
However, in this region, the largest component in the broadly spread components 
is outside the separatrix (see the yellow curve in 
Fig. \ref{fig:irep}(a)), meaning that the separatrix energy is 
contained in the non decaying region of $\bracket{J_\ell^{(M)}}{\Psi_0}$. 
Since $\bracket{q}{J_\ell^{(M)}}$ keeps constant when the position $q$ is outside 
the separatrix and the decaying speed of 
$\bracket{J_\ell^{(M)}}{\Psi_0}$ is so slow as shown in Fig. \ref{fig:irep_amp}, its product 
$\Con$
%$\bracket{q}{J_\ell^{(M)}}\bracket{J_\ell^{(M)}}{\Psi_0}$ 
also decays 
much slower than the instantion peak. 
Thus, at a certain critical $1/h_c$, the instanton component is overtaken 
by the dominant component in the broadly spread components. 
This is nothing but the I-NI transition \cite{ikeda_2013}. 

After the  I-NI transition, as long as the position of the dominant contribution 
in the broadly spread components is outside the separatrix, 
the decaying behavior in the plateau of $\bracket{J_\ell^{(M)}}{\Psi_0}$ controls 
the product $\bracket{q}{J_\ell^{(M)}}\bracket{J_\ell^{(M)}}{\Psi_0}$. 
This explains the presence of plateau in the splitting curve. 

However, note that the position of the dominant component is determined by 
the edge of the plateau of $\bracket{J_\ell^{(M)}}{\Psi_0}$, and 
this edge is located around the value $h/\tau$. 
As a result, at a certain value of $1/h$, the position of the dominant contribution 
passes through the separatrix (see Fig. \ref{fig:cont}(b)).  
If such an event occurs, the separatrix energy is then 
situated in the region where $\bracket{J_\ell^{(M)}}{\Psi_0}$ shows faster decay. 
This is exactly the moment when the splitting curve turns from the first plateau to 
the second steeply decaying region.

The mechanism generating the next plateau is understood by observing 
$\bracket{J_\ell^{(M)}}{\Psi_0}$ in a wider range. 
As shown in Fig. \ref{fig:action_rep}(a), a sequence of peaks appears at integer 
multiples of the fundamental energy unit $h/\tau$, and  
the decay rate of $\bracket{J_\ell^{(M)}}{\Psi_0}$ just below each resonance peak 
is again very slow as compared in the next sector, 
as demonstrated in Fig. \ref{fig:irep_amp}. Hence the same switching process takes place 
repeatedly.  We have actually checked that the mechanism explained here works at least 
until the third plateau, but we expect that this continues in larger $1/h$ regimes.

In this way, we could explain the emergence of the staircase structure 
based on the nature of the action representation, 
which seems to be closely connected with the fundamental energy sequence 
whose unit is given as $h/\tau$. 
As was checked above, the fundamental energy sequence can induce 
quantum resonances, resulting in the spikes in the splitting curve. 
However, it should be noted that the appearance of quantum resonances 
is not a necessary condition for the presence of the staircase structure, 
as discussed in the subsection \ref{subsec:tunnel_splitting} and \ref{subsec:I-NI}.
In other words, even if the resonance condition is not satisfied, 
a broadly spread or mild peak, whose width is almost comparable to 
the fundamental energy unit $h/\tau$, survives around the fundamental energy sequence. 
This is quite an anomalous situation because 
such a broad peak implies the existence of periodic oscillation of period $\tau$ 
accompanied by a rapid decaying process whose life time 
is comparable to the oscillation period itself \cite{ikeda_2014}. 

\begin{figure}[t]
	\centering
	\includegraphics[width=0.45\textwidth]{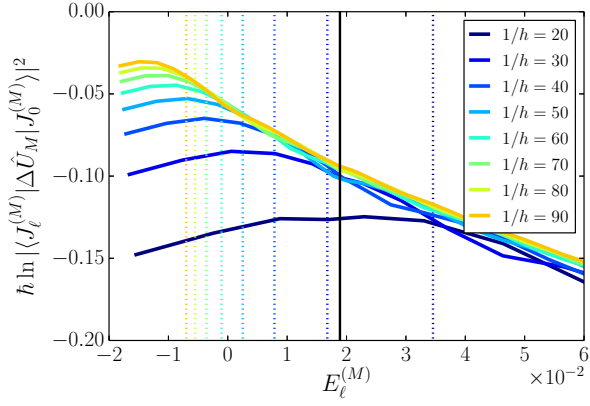}
	\caption{\label{fig:tunnel_scaling}
	(Color online)
Scaled wavefunction $\hbar \ln |\bra{J_\ell^{(M)}}\Delta U\ket{J_\ell^{(M)}}|^2$ as a function of $E_\ell^{(M)}$
for several effective Planck's constant $h$.
The black solid line and dotted lines respectively show
the separatrix energy, and the energies satisfying the condition $E=E_0^{(M)} + h/\tau$
}
\end{figure}

We also characterize this anomaly from the viewpoint of semiclassical theory. 
If the leading-order semiclassical approximation works, 
the matrix element $\bra{J^{(M)}_\ell}\Delta \hat{U}_M \ket{J^{(M)}_0}$ 
should take a form of $\Psi \sim \sum_\gamma A_{\gamma} e^{-iS_{\gamma}/\hbar}$, 
where $A_{\gamma}$ and $S_{\gamma}$ respectively stand for the amplitude and classical action, 
and the sum $\gamma$ is taken over complex classical orbits satisfying given initial and final conditions.
%associated with classical trajectory $\gamma$. 
In the semiclassical regime, we may neglect the $\hbar$ dependence in the amplitude $A_{\gamma}$, 
so the matrix element $\bra{J^{(M)}_\ell}\Delta \hat{U}_M \ket{J^{(M)}_0}$ 
is approximately expressed using the minimum imaginary action ${\rm Im}\,S_{\gamma_0}$
as $\Psi \sim e^{-{\rm Im}\,S_{\gamma_0}/\hbar}$. 
Since ${\rm Im}\,S_{\gamma_0}$ is a purely classical quantity, 
the form $\hbar\ln\bra{J^{(M)}_\ell}\Delta \hat{U}_M \ket{J^{(M)}_0}$ 
should not depend on $\hbar$. 
As will be shown in Fig. \ref{fig:scaled_irep_integ}, 
this is indeed the case in the integrable system. 
On the other hand, Fig. \ref{fig:tunnel_scaling} shows that 
the matrix element $\bra{J^{(M)}_\ell}\Delta \hat{U}_M \ket{J^{(M)}_0}$ 
does not follow the semiclassical ansatz in the non-decaying region, 
whereas the leading-order semiclassical prediction seems to work well 
beyond the non-decaying region. 
Although it is necessary to check whether or not the leading-order semiclassical approximation indeed breaks
in the non-decaying region,
the observed sharp distinction would be an important signature characterizing anomaly.

According to these speculations, 
we are currently taking two approaches to understand what was observed 
in the eigenstate  $\bracket{J_\ell^{(M)}}{\Psi_0}$ in the action representation; 
one is a real semiclassical analysis 
which is based on the so-called classical-quantum correspondence principle. 
This could extract anomalous components hidden in classical dynamics generated 
by the BCH Hamiltonian, and actually reproduce anomalous decay tails \cite{ikeda_2014}. 
Another approach is to take into account higher-order effects in the semiclassical analysis. 
Since 
similar non-decaying or anomalous behaviors 
have been found in the model with discontinuity in phase space, 
%the emergence of the non-decaying or anomalous regions in the action representation 
%invokes diffraction in phase space \cite{ishikawa_2012}, 
observed phenomena might be liked to or have at least close similarity 
with diffraction \cite{ishikawa_2012}.
This naturally leads us to the semiclassical treatment 
beyond the leading order.  
%We also note that analogous eigenfunctions 
%appear in a model which has been designed to invoke diffraction in phase space \cite{ishikawa_2012}. 
In any case, these are out of the scope of the present paper, and 
will be reported closely in our forthcoming paper \cite{hanada_2015}.

\section{Splitting curves in integrable systems}\label{sec:rat_integ}

\begin{figure}[t]
	\centering
	\includegraphics[width=0.45\textwidth]{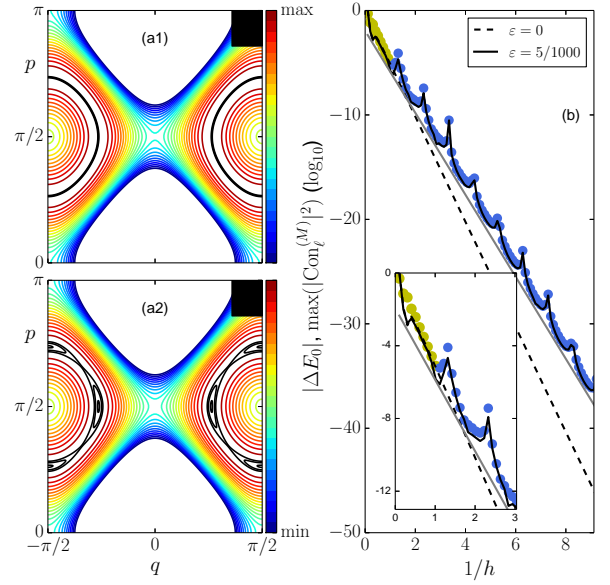}
	\caption{\label{fig:jeremy_normal}
	(Color online)
	Phase space portrait for the Hamiltonian (\ref{eq:jeremy_normal}) for $a=-0.55$ and 
	(a1) $\varepsilon=0$ and  (a2)  $\varepsilon=5/1000$.
	The black curves show the energy contour whose energy value is close to the maximum 
	one. The black box put in the upper right corner represents the size of 
	effective Planck's constant for 
	$h=1/5$. 
	(b) The splitting $\Delta E_0$ (in $\log_{10}$ scale) as a function of $1/h$ 
	in the cases of $\eps=0$ (black dashed line)
	and $\eps = 5/1000$ (black solid line).
	Yellow ad blue dots represent 
	the maximal mode of the contribution spectrum $\Con$ at $q=0$
	for $\ell=0$ and for $0 < \ell/N < 1/2$, respectively. 
	The gray 
	line shows the slope of the splitting curve for  $\eps=5/1000$.
	The inset is magnification of a small $1/h$ regime.
}
\end{figure}

In the previous subsection, we discussed the underlying mechanism 
controlling the staircase structure of the splitting curve and 
found that anomalous tails in eigenfunctions in the action representation play a key role.
If such a feature is shared only in nonintegrable maps, we would not expect 
the enhancement of the tunneling probability in the completely integrable system. 
%caused by the staircase 
Below we shall explain, 
the nature of the splitting curve in the integrable system 
is totally different, although a seemingly common behavior is observed.

\begin{figure}[t]
	\centering
	\includegraphics[width=0.49\textwidth]{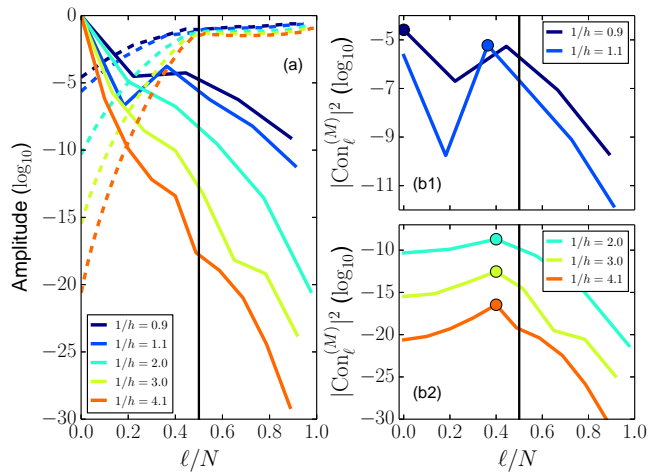}
	\caption{\label{fig:cont_and_irep}
	(Color online)
	(a) The amplitude of $\bracket{q}{\Psi_\ell}$ at $q=0$ (dashed curves) and 
	the action representation $\bracket{J_n}{\Psi_0}$ (solid curves)
	as a function of the normalized quantum number $\ell/N$.
	Right panels give the contribution spectrum $\Con$ ($\log_{10}$) as 
	a function of the normalized quantum number $\ell/N$ for (b1) a small $1/h$ regime,
	and (b2) a semiclassical regime.
}
\end{figure}

For this purpose, let us consider the following classically integrable Hamiltonian
\begin{equation}
  \label{eq:jeremy_normal}
  H(q,p) = H_0(q,p) + \eps H_1(q,p)
\end{equation}
with 
\begin{subequations}
  \begin{align}
    H_0(q,p) &= (\cos^2 q +\cos^2 p)/2+ a(\cos^2 q + \cos^2 p)^2,\\
    H_1(q,p) &= \cos^4 p - 6\cos^2 p \cos^2q + \cos^4 q.
  \end{align}
\end{subequations}
This system was analyzed in \cite{deunff_2013} to examine the validity of RAT theory 
in a completely integrable situation.
The authors have introduced a parameter $\phi$ which controls 
the relative orientation of the classical resonance chains \cite{deunff_2013}. 
Since the formulation of RAT theory do not take into account such orientation, 
RAT calculation could not follow the difference originating from it \cite{deunff_2013}. 

As is seen from the Figs. \ref{fig:jeremy_normal}(a1) and (a2), the equi-energy surface 
has a local maximum between an unstable fixed point $(q,p)=(0,0)$ and a stable fixed point 
$(q,p)=(\pm\frac{\pi}{2},0)$.  Some equi-energy surfaces in the inner well, which appear 
around the stable fixed points $(q,p)=(\pm\frac{\pi}{2},0)$, 
have the same energies as those in the outer region.  
For $\eps>0$, a classical nonlinear resonance chain is developed along the ridge between 
the inner well and outer region. 
%Note that the minimum energy are $(q,p)=(0,0)$ and $(-\pi,0)$ in this situation.

We impose the periodic boundary condition on the region 
$(q,p) \in (-\pi,\pi] \times (0,\pi]$,
%region $q\in(-\pi,\pi]$ and $p\in(0,\pi]$, 
and solve the eigenvalue problem
\begin{equation}
  \hat{H}(\hat{q},\hat{p})\ket{\Psi^{\pm}_n} = E_n^{\pm}\ket{\Psi_n^{\pm}}. 
\end{equation}
We then consider the splitting $\Delta E_0 = E_0^{+} - E_0^{-}$ of the ground 
and first exited states, both localizing in the inner well. 
Here we take the innermost state in the inner well as the ground state 
and arrange the eigenstates in the same order as the standard map.

Figure \ref{fig:jeremy_normal}(b) gives the splitting $\Delta E_0$ as a function of $1/h$.
For $\eps > 0$, the splitting decays exponentially accompanied with 
periodic spikes.  All the features have clearly been accounted for if one applies 
the semiclassical method using complex paths \cite{deunff_2013}.
The spikes appear as a result of the energetic resonance between 
the states localized in the inner well and outer region. The coupling strength 
could be evaluated using the imaginary action of complex trajectories 
which bridge classical disjointed equi-energy surfaces. 

It would be worth mentioning that for $\eps=0$ the condition $H(q,p) = 0$ 
can be factorized into 
\begin{equation} \label{eq:ecurve1}
  \cos^2q + \cos^2p =0,
\end{equation}
and 
\begin{equation}\label{eq:ecurve2}
  \cos^2q + \cos^2p = -1/2a.
\end{equation}
This shows that the invariant curves specified by (\ref{eq:ecurve1}) and (\ref{eq:ecurve2}) 
are not connected even in the complex plane, thus 
no tunneling connection between the inner and outer regions exists 
even though both are the surfaces with the same energy \cite{Harada}. 
As a result, the splitting $\Delta E_0$ 
exhibits single exponential decay without spikes.

On the other hand, with careful observation of the splitting curve for $\eps >0$ 
(see Fig. \ref{fig:jeremy_normal}(b)), we notice that there exists a crossover 
from one slope to another. 
In a small $1/h$ regime, the slope can be well fitted by the one for $\eps=0$, 
whereas the best fit curve, colored in gray in Fig. \ref{fig:jeremy_normal}(b), shows 
another slope in the large $1/h$ regime.

Such a crossover or the change of the slope of the splitting curve 
reminds us of the plateau discussed in the nonintegrable situation.
However, the origin and the underlying mechanism entirely 
differs from the previous one.
%, as explained. 
This can be confirmed again by examining the contribution spectrum. 
Here we use the eigenstates $\ket{J_n}$ as the basis states
for the contribution spectrum, 
where $\hat{H}_0(\hat{q},\hat{p})\ket{J_n} = E_n^{(0)}\ket{J_n}$.

\begin{figure}[t]
	\centering
	\includegraphics[width=0.45\textwidth]{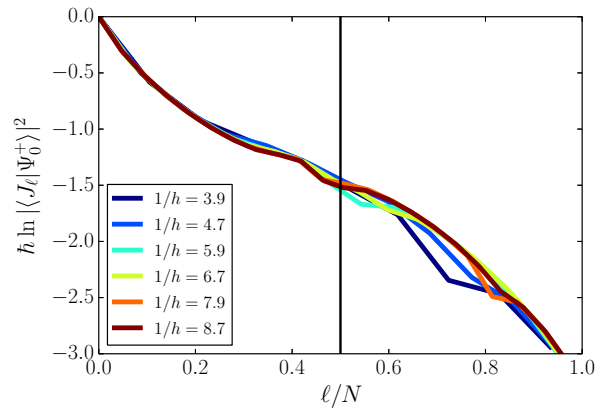}
	\caption{\label{fig:scaled_irep_integ}
	(Color online)
Scaled wavefunciton $\hbar \ln |\bracket{J_\ell}{\Psi_0^+}|^2$ as a function $\ell/N$ 
for several effective Planck's constant $h$, 
each of which is taken at the values of off resonance positions in Fig.\ref{fig:cont_and_irep}(b).
%each of which 
%is taken at the values off the resonance positions in Fig. \ref{fig:cont_and_irep}(b).
}
\end{figure}

\begin{figure}[t]
	\centering
	\includegraphics[width=0.5\textwidth]{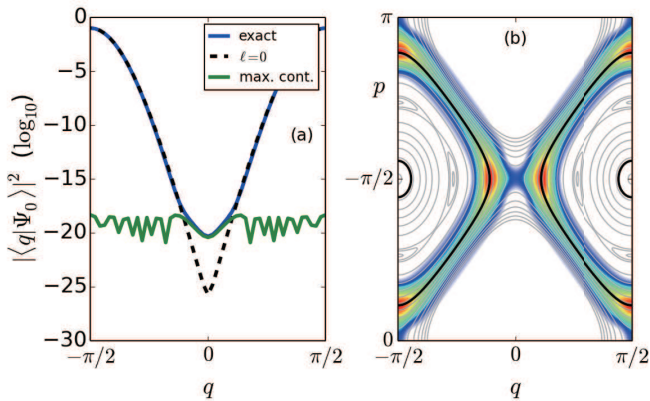}
	\caption{\label{fig:jeremy_hsm}
	(Color online)
	(a) The eigenfunction $\ket{\Psi_0}$ in the $q$-representation for $\eps = 5/1000$ (blue), 
	the eigenfunction $\ket{J_0}$ (dotted) and the maximal mode in the contribution spectrum
	at $q=0$ (green), respectively. 
	(b) The maximal mode state in the Husimi representation for $h=1/5$. 
	The black thick curves represent inner and outer invariant curves with 
	the same energy $E=E_0$. 
}
\end{figure}

As shown in Fig. \ref{fig:cont_and_irep}(b1), a switching from the instanton  
to another mode also occurs, like the standard map case. 
However, in the integrable case, 
the position of the peak sits at the same value of $\ell/N$ and does not move 
even if the value of $1/h$ is changed, while remember that it depends on $1/h$ and shift leftwards 
in case of the standard map.
Note here that $\ell/N$ can be identified with the action coordinate. 
The reason for the peak position being fixed is simple; the peak appears as a result of 
the coupling between inner and outer surface, which is expected to occur in the 
RAT scenario. 
Alternatively stated, the origin of coupling is purely classical. 
Furthermore, as shown in Fig. \ref{fig:scaled_irep_integ}, 
the leading-order semiclassical anstaz, which was discussed in the previous section, 
works quite well for the eigenfunction in the action representation.
These results make a sharp contrast to the standard map case.
%This makes a sharp contrast to the standard map case. 
We can see in Fig. \ref{fig:jeremy_hsm} that the maximal mode in the contribution spectrum 
well reproduces the structure of eigenfunction around $q=0$, and its support is exactly 
an invariant curve with the same energy as that of the ground state.

The presence of the crossover admits a simple semiclassical interpretation. 
As discussed in \cite{deunff_2013}, there exist two different complex paths 
with different imaginary actions.  One corresponds to the ordinary instanton path, 
which runs from one well to another directly and the other is the path bypassing the 
classical resonance chain. 
In the semiclassical regime, since the latter one has a smaller imaginary action. 
On the other hand, in a small $\eps$ regime, it can happen that the instanton contribution 
is larger than that from the bypassing one, in spite of the magnitude relation of 
imaginary actions. 
This is because the prefactor, more precisely 
the coupling amplitude due to tunneling, comes into play in a relatively small $1/h$ regime. 
The observed crossover would be understood 
by taking into account not only the imaginary action but the coupling amplitude. 
%%%%% EndEndEnd %%%%%%
%Such a situation is indeed realized in Fig. \ref{fig:cont_and_irep}(b1). 
This argument suggests that, in a larger $\eps$ regime, the coupling with bypassing path 
gets larger, and the crossover point disappears when the value of $\eps$ exceeds a 
certain threshold. 
Note, however that 
%This is exactly the situation we observe in Fig. \ref{fig:cont_and_irep}(b1). 
%However, 
the splitting curve cannot form the staircase structure since we have at most two possible 
complex paths, and the underlying mechanism generating spikes has a purely classical origin 
as stated above. 
%This argument makes sense as far as the semiclassical approximation works.

%%%%%%%%%%%%%%%%%%%%%%%%%%%%%%%%%%%%%%%%%%%%%%%%
%
\section{Summary and outlook}
\label{sec:summary}

The focus of the present paper was put on clarifying the origin of the enhancement 
of tunneling probability in the nearly integrable system. 
We here measured the tunneling probability by observing tunneling splittings plotted 
as a function of the inverse Planck's constant. 
Typical features of the splitting curve commonly observed in nonintegrable quantum maps 
are the existence of spikes and persistent departure of the splitting curve from the one 
predicted by the instanton. 

So far these have been discussed in the framework of RAT theory, but here we took 
a different perspective: the splitting curve is composed of 
the staircase-shaped backbone accompanied by spikes. 
We have observed that, by introducing the absorber composed of integrable bases, 
spikes could be selectively suppressed if states interacting with the reference doublet are
absorbed.  More precisely, we have shown that the interacting or third state could be decoupled from 
the reference doublet when we take an integrable state that 
has maximally overlap with the corresponding eigenstate as the absorber.  
The eigenenergy of such eigenstates responsible for generating spikes are 
pushed out to the complex plane, and spikes disappear. 
Our observation was that even though all the third states which resonate energetically 
with the reference doublet are suppressed in such a way the staircase structure survives. 
Note that the efficiency of the present absorber comes from the fact that 
the regime we consider is close enough to the integrable limit, otherwise 
the absorber may affect irrelevant states in an uncontrollable way. 

The result strongly suggests the existence of non-trivial broad interaction between the reference doublet 
and other states. This was indeed confirmed by introducing renormalized Hamiltonian, 
which is constructed using the BCH expansion, and used as basis states by which the 
reference state is expanded. 
In particular, we focus on the contribution spectrum at the origin $q=0$ since the amplitude 
of eigenfunctions at the origin follows the behavior of tunneling splitting. 
Here the contribution spectrum introduced in \cite{ikeda_2013} represents 
components of renormalized states in the reference state. 

The contribution spectrum analysis clearly revealed that, in addition to the self component 
representing the instanton, 
there certainly exists broad interaction, and the behavior of such broadly spread components 
controls the staircase structure in the splitting curve. 
There are two key ingredients to explain the emergence of the staircase: 
one is the behavior of the most dominant state in broad components, 
the other is anomalous tail observed in the eigenfunction in the action representation. 
Note that the renormalized bases are crucially important to capture these features, 
otherwise one could not explain the existence of the staircase structure
and the anomalous tail part in the action representation as well. 

The dominant contributor in the broadly spread components switches from one to another, 
which was observed in the contribution spectrum. 
Such a switching phenomenon is driven by and liked to 
the existence of the fundamental energy sequence, 
which is further enhanced when the quantum resonance 
between unperturbed system and the periodic driving occurs.
%quantum resonance which appears as  
%a result of the resonance between unperturbed system and the periodic driving force. 
%This type of resonances occurs even though the state interacting the reference states 
%is sitting outside the inner torus region. The relation between the resonance in this sense 
%and the one discussed in RAT theory is not clear enough at the moment.  

The origin of anomalous part in the action representation should be 
explored more closely, which will become a primary subject of our forthcoming papers. 
The semiclassical analysis based on the correspondence principle, in which 
not complex but real classical orbits are used as input information. 
This efficiently works and turns out to extract anomalous components in classical dynamics 
of the BCH Hamiltonian \cite{ikeda_2014}. 
The analogy with the system modeling the diffraction, 
together with some speculations on anomalous behaviors of caustics appearing 
in the semiclassical analysis will be another approach \cite{hanada_2015}. 
The latter suggests that observed phenomena in the eigenfunction in the action 
representation are beyond the leading semiclassical description.  
%
%, tells us that this region is beyond the leading semiclassical 
%description.  Such an argument is consistent with the observed fact that the decay rate 
%in the anomalous part is prominently different from that for the outer part.  

These two key characteristics are, by their very nature, absent in the completely integrable system. 
Therefore, one could predict that the staircase structure does not appear in the
completely integrable. We have confirmed this for a normal form Hamiltonian system, 
for which the validity of RAT theory was recently investigated. 
We have shown that a sharp contrast exists between integrable and nonintegrable systems 
and verified that the dominant contributor in the contribution spectrum for the 
integrable system sits at the same position and does not move as in the nonintegrable case. 
The absence of the staircase structure could simply be interpreted by the fact that 
there exists a unique dominant complex path in the semiclassical regime. 

Finally we would like to emphasize the importance of observing 
wavefunctions in the whole range, not focusing only on the amplitude at a specific point, 
like the origin $q=0$ in the present case. 
%although it is correlated with tunneling splitting, or on the tunneling rate measured 
%at a certain specific point. 
As discussed in subsection \ref{subsec:I-NI}, with increase in $1/h$, the convex structure around the 
origin, appearing in the first plateau, is pushed outward and forms shoulders in both sides. 
The same process happens repeatedly as one further increases $1/h$,  that is, similar shoulders 
appear one after another. In this sense, we can find the trace of the staircase of the 
splitting curve in the tail pattern of wavefunction.  
This is also true for wavefunction in the action representation. 
There exists a significant difference between inner and outer tunneling tail, and this exactly results in 
different slopes of the splitting curve and thus staircase skeleton.

\section*{ACKNOWLEDGMENTS}
We are grateful for useful discussions with 
H. Harada, J.  Le Deunff, A.  Mouchet, T. Okushima, and K. Takahashi.  
We especially thanks to N. Mertig for his helpful comments on RAT theory. 
This work has been supported by JSPS KAKENHI Grant
Numbers 24340094 and 25400405. 
The authors appreciate Shoji Tsuji and Kankikai 
for using their facilities at Kawaraya during this study.

%\bibliography{cite}

\begin{thebibliography}{99}
\bibitem{saddle_numbering}
 In the vicinity of the energy corresponding to the unstable fixed point at 
 $(q,p)= (0,0)$, the states $\ket{\Psi_n^{\pm}}$ do not form tunneling doublets, so 
 this numbering rule does not work, but this ambiguity will not cause 
 any serious confusion in the present description.
\end{thebibliography}
%\nocite{*}
\end{document}